\documentclass{pasj00}
\SetRunningHead{Takamori et al.}{Numerical Method for the Stationary Pulsar Magnetosphere}
\Received{}
\Accepted{}
\Published{}
\begin{document}
\title{An Alternative Numerical Method for the Stationary Pulsar Magnetosphere} 
\author{Yohsuke \textsc{Takamori}}
\affil{Osaka City University Advanced Mathematical Institute (OCAMI),
3-3-138 Sugimoto, Sumiyoshi, Osaka 558-8585, Japan}
\email{takamori@sci.osaka-cu.ac.jp; takamori@wakayama-nct.ac.jp}
\author{Hirotada \textsc{Okawa}}
\affil{CENTRA, Departamento de F\'{\i}sica, Instituto Superior T\'ecnico, 
Universidade T\'ecnica de Lisboa - UTL, Av. Rovisco Pais 1, 1049 Lisboa, Portugal}
\email{hirotada.okawa@ist.utl.pt}
\author{Makoto \textsc{Takamoto}}
\affil{Max-Planck-Institut f$\ddot{u}$r Kernphysik, Saupfercheckweg 1, 69117 Heidelberg, Germany}
\email{makoto.takamoto@mpi-hd.mpg.de}
\and
\author{Yudai \textsc{Suwa}}
\affil{Yukawa Institute for Theoretical Physics, Kyoto University, Oiwake-cho, Kitashirakawa, Sakyo-ku, Kyoto 606-8502, Japan}
\email{suwa@yukawa.kyoto-u.ac.jp}


\maketitle 

\begin{abstract}
Stationary pulsar magnetospheres in the force-free system are governed by the pulsar equation. 
In 1999, Contopoulos, Kazanas, and Fendt (hereafter CKF) numerically solved 
the pulsar equation and obtained a pulsar magnetosphere model called the CKF solution 
that has both closed and open magnetic field lines. The CKF solution is a successful solution, 
but it contains a poloidal current sheet that flows along the last open field line.
This current sheet is artificially added to make the current system closed.
In this paper, we suggest an alternative method to solve the pulsar equation and construct pulsar magnetosphere
models without a current sheet. In our method, the pulsar equation is decomposed
into Ampere's law and the force-free condition. We numerically solve these equations simultaneously 
with a fixed poloidal current. As a result, we obtain a pulsar magnetosphere model
without a current sheet, which is similar to the CKF solution near the neutron star and has a jet-like structure 
at a distance along the pole. In addition, we discuss physical properties of the model
and find that the force-free condition breaks down in a vicinity of the light cylinder due to
dissipation that is included implicitly in the  numerical method.
\end{abstract}

\section{Introduction}
Pulsar magnetospheres play an important role on activities of the pulsars. 
For example, the magnetic field can extract  the rotational energy and convert it into radiation. 
Thanks to strong electromagnetic field, charged particles can be accelerated so as to form a pulsar wind. 
In addition, because of the acceleration, photons are emitted from those charged particles, 
which becomes a source of X-rays or gamma-rays. Thus in order to understand pulsar physics, 
it is important to study the magnetic field structure around pulsars.

Studies of pulsar magnetospheres started in the context of electromagnetostatics in vacuum.
The simplest model of the pulsar magnetosphere is that of a dipole, with the power of radiation from
the pulsar explained as dipole radiation \citep{1967Natur.216..567P}. 
However, \citet{1969ApJ...157..869G} pointed out that because of its rapid rotation in the strong magnetic field, 
the surroundings of the neutron star should be filled with plasma, so that study of the pulsar magnetosphere 
must take account of the motion of the plasma.

The system becomes highly non-linear, and therefore it is difficult to determine self-consistently 
both the motion of the plasma and the configuration of the energy of the electromagnetic field, 
the inertia of the plasma is ignored. Because pulsars have strong magnetic fields, 
this assumption would be sufficiently satisfied. 
Moreover, when the system is stationary and axisymmetric, the electromagnetic field is represented by three quantities:
the magnetic flux $\Psi$, the poloidal current $I_{\rm P}(\Psi)$, and the angular velocity $\Omega_{\rm F}(\Psi)$
of the magnetic field line. The pulsar magnetosphere in the force-free system is determined by the pulsar
equation \citep{1973ApJ...180..207M,1973ApJ...182..951S,1973Ap&SS..24..289M,1974MNRAS.167..4570}
which is a quasi-linear elliptic-type differential equation for $\Psi$ with $I_{\rm P}(\Psi)$ and $\Omega_{\rm F}(\Psi)$ as its source terms. 
For the zero-poloidal current, a solution with dipole geometry has been constructed \citep{1979MNRAS.188..799M,1973ApJ...180L.133M}. 
Some cases with a non-zero poloidal current have been also discussed 
\citep{1973ApJ...180..207M, 1973ApJ...180L.133M, 1976MNRAS.176..465B, 
1983ZhETF..85..401B, 1990ApJ...350..732S, 1990SvAL...16...16L,
1995A&A...300..791F, 1998MNRAS.298..847B}.
Although they could construct a pulsar mangetosphere filled with plasma, 
due to the singularity of the pulsar equation at the so-called light cylinder,
some solutions can be applied only inside the light cylinder. The others pass through the light
cylinder smoothly, but it is hard to apply them to realistically astrophysical situation.

In order to construct realistic pulsar magnetosphere models across the light cylinder,
\citet{1999ApJ...511..351C} suggested a new numerical method which is called the CKF
method and obtained a solution, the so-called CKF solution which has a dipole geometry near the
neutron star and passes through the light cylinder smoothly. After their work, other properties of the
CKF solution (e.g., the drift velocity, the energy loss rate) have been discussed 
\citep{2003PThPh.109..6190, 2004MNRAS.349..213G, 2005PhRvL..94b1101G, 2006MNRAS.368.1055T}.
Interestingly, dynamical simulations of force-free electrodynamics or magnetohydrodynamics have shown 
that a CKF-like configuration appears as the steady state \citep{2006MNRAS.367...19K, 2006MNRAS.368L..30M, 2006ApJ...648L..51S}.
Moreover, the CKF method was extended to construct magnetospheric models 
in the more general case with the angular velocity of the magnetic field line $\Omega_{\rm F}$
being not constant \citep{2005A&A...442..579C, 2007Ap&SS.308..575T}.

The CKF method is a successful historical milestone, and it has been used to study pulsar physics. 
However, there still remains a problem. The obtained solution
by the CKF method has a poloidal current sheet that flows along the equator and the last open field
line (hereafter, we simply call it ``current sheet''). This current
sheet is artificially added to make the current system closed. 
\citet{2003ApJ...598..446U} focused on a local region near the Y-point, which is the
point where the current sheet splits in two currents, and imposed the
condition that the Maxwell stress at the last open field line be
continuous. It was subsequently pointed out that the electromagnetic field 
becomes infinite if the Y-point is located on the light-cylinder; in other words, if one admits the current sheet, the Y-point should locate
inside the light cylinder to avoid this infinity. 
Such cases have been studied in \citet{2004MNRAS.349..213G} and \citet{2006MNRAS.368.1055T}, 
and they obtained solutions that pass smoothly through the light cylinder. However, the electromagnetic field has the minimum energy in the
case that the Y-point locates exactly on the light cylinder \citep{2006MNRAS.368.1055T} so that the solution with
the minimum energy is unphysical because of Uzdensky's result. To sum up, it is worth while constructing
a magnetosphere model ``without'' the introduction of a current sheet such as that in the CKF solution.

The problem was taken up by \citet{2006ApJ...652.1494L}.
Although they constructed an alternative model of the pulsar magnetosphere
without a current sheet that has a jet flow and a disk wind near the equator, they did not
construct a CKF-type structure without a current sheet. Meanwhile
\citet{2003PThPh.109..6190} tried to construct a CKF-type structure
without a current sheet, 
but they could not obtain such solutions. 
In this paper, we solve the pulsar equation and construct a CKF-type structure without a current sheet.
It seems that the CKF method cannot be used for constructing a CKF-type structure without a current
sheet \citep{2003PThPh.109..6190}.
We therefore suggest an alternative numerical method to construct pulsar magnetosphere models with fixed $I_{\rm P}(\Psi)$.
Moreover, developing an alternative numerical method is worth while from a mathematical point of view
because there is no mathematical rigorous proof of the existence of a solution 
which is smooth over the light cylinder for the pulsar equation. 
This fact arouses our suspicion 
about a numerical solution of the pulsar equation, 
that is, whether the numerical solution really represents a solution of the pulsar equation or not.
Although the mathematical proof is too difficult, we can expect to get some information 
about physical meanings of the numerical solution by looking the pulsar equation from various angles.

This paper is organized as follows. In \S 2, we summarize the derivation of the pulsar equation
and introduce the CKF method. In \S 3, we explain our method to solve the pulsar equation with fixed
$I_{\rm P}(\Psi)$ and study pulsar magnetosphere models both with and without a current sheet. 
Then, we also study physical properties of our results: the energy loss rate; the three-dimensional magnetic field
structure in \S 4. Finally, \S 5 is devoted to summary and discussion.

Through this paper, we use Gaussian units.
\section{CKF method}\label{CKF}
In this section, we briefly summarize the derivation of the pulsar equation and
the numerical procedure, the so-called CKF method, 
to solve the pulsar equation suggested by \citet{1999ApJ...511..351C}.

Throughout this paper, we focus on stationary and axisymmetric
electromagnetic fields in the force-free system. Let $\vec{E}$ and
$\vec{B}$ be the electric field and the magnetic field, respectively.
Maxwell's equations with stationarity are given by
\begin{eqnarray}
 \vec{\nabla}\cdot\vec{B} &=& 0, \\
 \vec{\nabla}\times\vec{E} &=& 0, \\
 \vec{\nabla}\cdot\vec{E} &=& 4\pi\rho_e, \\
 \vec{\nabla}\times\vec{B} &=& \frac{4\pi}{c}\vec{J},
\end{eqnarray}
where $\rho_e$ and $\vec{J}$ are the electric charge density and the
electric current density, respectively, and $c$ is the speed of light.
In addition, the force-free condition is given by
\begin{equation}
 \rho_e\vec{E}+\frac{1}{c}\vec{J}\times\vec{B}=0.
 \label{FF}
\end{equation}
Thanks to the symmetries and
the force-free condition, the electric and the magnetic field can be
expressed by the following three physical quantities; the
magnetic flux $\Psi$, the angular velocity of the magnetic field lines
$\Omega_{\rm F}$, and the poloidal electric current $I_{\rm P}$. 
From the force-free condition, we can see that
$\Omega_{\rm F}$ and $I_{\rm P}$ are arbitrary functions of $\Psi$ as follows.

First, let us define the magnetic flux $\Psi$ and the electric current $I_{\rm P}$.
These are determined by the following surface integral;
\begin{eqnarray}
 \Psi &:=& \int_{\cal A} \vec{B}\cdot d\vec{S}, \\
 I_{\rm P} &:=& \int_{\cal A} \vec{J}\cdot d\vec{S}, 
\end{eqnarray}
where $\cal A$ is an axisymmetric two dimensional spacelike surface.
Using these functions and the poloidal components of Ampere's law, in 
cylindrical coordinates $(R,\varphi,z)$, the magnetic field $\vec{B}$ can be written as
\begin{equation}
 \vec{B} = \frac{\vec{\nabla}\Psi\times\vec{e}_{\varphi}}{2\pi R}
 +\frac{2I_{\rm P}(R,z)}{cR}\vec{e}_{\varphi},
 \label{B}  
\end{equation}
where $\vec{e}_{\varphi}$ is a unit vector tangent to the $\varphi$ direction.
Furthermore, the stationary electric fields can be expressed 
by an electric potential $\Phi(R,z)$ as 
\begin{equation}
 \vec{E}=-\vec{\nabla}\Phi.
 \label{E}
\end{equation}
Thus, in general, stationary and axisymmetric electromagnetic fields can be expressed
by $\Psi(R,z)$, $I_{\rm P}(R,z)$, and $\Phi(R,z)$.

In addition, by imposing the force-free condition, we find that
$I_P$ and $\Omega_{\rm F}$ become functions of $\Psi$.
Taking the inner product of Eq.~(\ref{FF}) with $\vec{B}$, we have
\begin{equation}
 \vec{E}\cdot\vec{B}=0.
\end{equation}
Substituting Eqs.~(\ref{B}) and (\ref{E}) into the above equation, 
we find that the electric potential $\Phi$ becomes a function of $\Psi$,
that is, $\Phi=\Phi(\Psi)$. Therefore, the electric field can also be expressed by $\Psi$.

Next, let us consider the toroidal component of the force-free condition. We have
\begin{equation}
 \frac{1}{c}\vec{J}_{\rm P}\times\vec{B}_{\rm P}=0,
 \label{FF:polo}
\end{equation}
where the subscript P means the poloidal components. 
From the definition of $I_{\rm P}$, we have
\begin{equation}
 \vec{J}_{\rm P}=\frac{\vec{\nabla}I_{\rm P}\times \vec{e}_{\varphi}}{2\pi R}.
 \label{Jp}
\end{equation}
Thus, Eq.~(\ref{FF:polo}) gives
\begin{equation}
 (\partial_R I_{\rm P})\partial_z\Psi-(\partial_z I_{\rm P})\partial_R\Psi=0,
\end{equation}
which implies that $I_{\rm P}$ is a function of $\Psi$. 

Finally, the electric field $\vec{E}$ and the magnetic field $\vec{B}$ can 
be rewritten in the following form;
\begin{eqnarray}
 \vec{E} &=& -\frac{\Omega_{\rm F}(\Psi)}{2\pi c}\vec{\nabla}\Psi, \label{E_ff}\\
 \vec{B} &=& \frac{\vec{\nabla}\Psi\times\vec{e}_{\varphi}}{2\pi R}+
 \frac{2I_{\rm P}(\Psi)}{cR}\vec{e}_{\varphi}
 \label{B_ff},
\end{eqnarray}
where we define $\Omega_{\rm F}(\Psi):=2\pi c d\Phi/d\Psi$. 
Thus a stationary and axisymmetric electromagnetic field
with the force-free condition can be expressed by three scalar functions  $\Psi$, $\Omega_{\rm F}(\Psi)$, and $I(\Psi)$.
Substituting Eqs.~(\ref{E_ff}) and (\ref{B_ff}) into the 
Gauss's law and Ampere's law,  we  have
\begin{eqnarray}
 -\vec{\nabla}\cdot{}\left(\frac{\Omega_{\rm F}(\Psi)\vec{\nabla}\Psi}{2\pi c}\right) &=& 4\pi\rho_e,
 \label{Gauss}\\
 -R\vec{\nabla}\cdot{}\left(\frac{\vec{\nabla}\Psi}{2\pi R^2}\right) &=& \frac{4\pi}{c} J_{\rm T},
 \label{Ampere}
\end{eqnarray}
where $J_{\rm T}:=\vec{J}\cdot{}\vec{e}_{\varphi}$.  
Moreover, we have left the poloidal components of the force-free condition.
From the poloidal components of Eq.~(\ref{FF}), 
\begin{equation}
  \rho_e \vec{E}+\frac{1}{c}\left(\vec{J}_{\rm P}\times\vec{B}_{\rm T}+\vec{J}_{\rm T}\times\vec{B}_{\rm P}\right)=0,
\end{equation}
whence from Eqs.~(\ref{Jp}), (\ref{E_ff}) and (\ref{B_ff}), we obtain the following equation,
\begin{equation}
 R\Omega_{\rm F}\rho_{e}+\frac{2I_{\rm P}}{cR}\frac{dI_{\rm P}}{d\Psi}=J_{\rm T}.
 \label{FFT}
\end{equation}
Combining Eqs.~(\ref{Gauss}), (\ref{Ampere}), and (\ref{FFT}), we find an
elliptic type differential equation for $\Psi$ as follows:
\begin{eqnarray}
 &&\vec{\nabla}\cdot{}\left(R^{-2}\left(1-\frac{R^2\Omega_{\rm F}^2}{c^2}\right)\vec{\nabla}\Psi\right)
 \nonumber \\
 ~~~~~&&+\frac{\Omega_{\rm F}}{c^2}\frac{d\Omega_{\rm F}}{d\Psi}\vec{\nabla}\Psi\cdot{}\vec{\nabla}\Psi
 +\frac{16\pi^2}{c^2R^2}I_{\rm P}\frac{dI_{\rm P}}{d\Psi}=0,
\end{eqnarray}
This equation is known as the pulsar equation
or the Grad-Shafranov equation in the limit of strong magnetic field. 
For pulsar magnetospheres, the angular velocity of the magnetic field lines,
$\Omega_{\rm F}$, is usually assumed as a constant because the
magnetic field lines penetrating the pulsar surface would co-rotate
rigidly with the pulsar. 
Even if the force-free approximation breaks down locally due to 
a kinematic effect of plasma, the condition $\Omega_{\rm F}$ that is function of $\Psi$
remains valid in the case of highly conducting plasma in general. 
For the pulsar case, electron-positron pair creation would occur 
in a polar region near the neutron star surface and they are accelerated by the electric field parallel to the magnetic field line.
The force-free approximation is not satisfied in the acceleration-region but the assumption 
that $\Omega_{\rm F}$ is a function of $\Psi$ remains valid in the region. 
The case of $\Omega_{\rm F}$ being not constant has been studied in \citet{2005A&A...442..579C} and \citet{2007Ap&SS.308..575T}.
Here, we assume that $\Omega_{\rm F}$ is a constant.
In this case, the pulsar equation becomes
\begin{eqnarray}
 &&\left(1-\frac{R^2\Omega_{\rm F}^2}{c^2}\right)\left(\partial_R^2\Psi-\frac{1}{R}\partial_R\Psi+\partial_z^2\Psi\right) \nonumber \\
 &&~~~~~-\frac{2R\Omega_{\rm F}^2}{c^2}\partial_R\Psi+\frac{16\pi^2I_{\rm P}}{c^2}\frac{dI_{\rm P}}{d\Psi}=0.
 \label{eq:pulsar}
\end{eqnarray}
Giving a functional form of $I_{\rm P}(\Psi)$ and imposing boundary conditions, 
we can obtain a solution of the pulsar equation. However, it is well known
that a smooth solution of the pulsar equation is difficult to obtain
due to the existence of the {\it light cylinder}.

It is easy to see that the surface where $1-R^2\Omega_{\rm F}^2/c^2=0$,
the so-called light cylinder, is singular in the pulsar equation.  
In order to make the derivatives of $\Psi$ finite at the light cylinder,
$R=R_{\rm LC}$, we impose one constraint (cf. (\ref{eq:pulsar}))
\begin{equation}
 -2R_{\rm LC}\Omega_{\rm F}^2\partial_R\Psi|_{R=R_{\rm LC}}
+16\pi^2I_{\rm P}\frac{dI_{\rm P}}{d\Psi}\bigg{|}_{R=R_{\rm LC}}=0.
 \label{LC}
\end{equation}
If we fix the functional form of $I_{\rm P}(\Psi)$, Eq.~(\ref{LC}) becomes the
Neumann boundary condition at the light cylinder, and hence the pulsar
equation should be solved both inside and outside the light cylinder
independently. It implies that the solutions would not be smooth at
the light cylinder in general.  

\citet{1999ApJ...511..351C} pointed out that Eq.~(\ref{LC}) can 
be regarded as an equation to determine the functional form of $I_{\rm P}(\Psi)$.
In order to obtain a smooth solution beyond the light cylinder,
they developed the following numerical procedure, the so-called CKF method:
\begin{enumerate}
 \item Choose a trial poloidal electric current $I_{\rm P}(\Psi)$.
 \item By solving the pulsar equation with the trial $I_{\rm P}$ both inside ($-$) and outside ($+$) the light cylinder,
       two magnetic flux functions, $\Psi_{\mp}(R,z)$, are obtained. 
       In general, the value of $\Psi_{-}$ differs from that of $\Psi_{+}$ at the light cylinder.
 \item By substituting $\Psi_{\pm}$ into Eq.~(\ref{LC}), respectively, two different poloidal electric currents,
       $I_{\rm P\pm}dI_{\rm P\pm}/d\Psi$, are obtained at the light cylinder. 
       Then, a new poloidal electric current is determined by
       \begin{eqnarray}
 	&&16\pi^2I_{\rm P}\frac{dI_{\rm P}}{d\Psi}(\Psi) \nonumber \\
	 &&= \mu_116\pi^2I_{\rm P+}\frac{dI_{\rm P +}}{d\Psi}(\Psi_{+})
	 +\mu_216\pi^2I_{\rm P-}\frac{dI_{\rm P -}}{d\Psi}(\Psi_{-}) \nonumber \\
	 &&~~+\mu_3(\Psi_+-\Psi_-),
       \end{eqnarray}
       for
       \begin{equation}
	\Psi=\frac{1}{2}(\Psi_++\Psi_-),
       \end{equation}
       where $\mu_1$, $\mu_2$, and $\mu_3$ are weight factors satisfying $\mu_1+\mu_2=1$ and $\mu_3\ll 1$.
 \item Repeat these steps until the difference of $\Psi_\pm$ becomes numerically negligible at the light cylinder.
\end{enumerate}
Although there is no guarantee whether this scheme converges, 
they succeeded in obtaining so-called the CKF solution.
In the CKF method, since $I_{\rm P}dI_{\rm P}/d\Psi$ changes in the step 3, the functional form of
$I_{\rm P}(\Psi)$ cannot be determined until the calculation is finished. 
Note that one should integrate $I_{\rm P}dI_{\rm P}/d\Psi$ to obtain $I_{\rm P}(\Psi)$. 
At the z-axis, the electric current should satisfy $I_{\rm P}=0$ because there is no electric current there. 
Moreover, $I_{\rm P}$ also should be zero at the last open field line to make the current system closed.
However, since the equation for $I_{\rm P}$ is a first order differential equation, we can impose only
one boundary condition. For the CKF solution, the boundary condition $I_{\rm P}=0$ at the axis is imposed.
As a result, a current sheet flowing along the last open field line is added artificially to 
make the current system closed.

One of our purposes is to construct a pulsar magnetosphere model 
without a current sheet along the last open field as introduced in the CKF solution. 
\citet{2003PThPh.109..6190} imposed no current sheet condition in the 
CKF method and carried out their calculation, but they could not obtain such solutions. 
Thus, it seems that the CKF method is not useful for obtaining solutions without a current sheet.
In order that, we want to study the pulsar equation with $I_{\rm P}(\Psi)$ fixed.
In the next section, we will suggest an alternative numerical method so as to obtain 
a solution without a current sheet and show results. 

\section{Numerical Study of Pulsar Magnetospheres with Fixed $I_{\rm P}(\Psi)$}\label{our_method}
\subsection{Numerical procedure}
In order to study the pulsar equation with fixed $I_{\rm P}(\Psi)$, 
let us decompose the pulsar equation, (\ref{eq:pulsar}), into Ampere's law and the 
force-free condition as follows:
\begin{eqnarray}
 &&\partial_R^2\Psi-\frac{1}{R}\partial_R\Psi+\partial_z^2\Psi=-\frac{8\pi^2}{c}RJ_{\rm T}(R,z), \label{Ampere2} \\
 &&\frac{R^2\Omega_{\rm F}^2}{c^2}\left(\partial_R^2\Psi-\frac{1}{R}\partial_R\Psi+\partial_z^2\Psi\right)
 \nonumber\\ 
 &&~~+\frac{2R\Omega_{\rm F}^2}{c^2}\partial_R\Psi-\frac{16\pi^2 }{c^2}I_{\rm P}I_{\rm P}'
=-\frac{8\pi^2}{c}RJ_{\rm T}(R,z)
 \label{FF2}.
\end{eqnarray}
where the prime denotes the derivative with respect to $\Psi$.
The first equation is Ampere's law and the second equation is the force-free condition with Gauss's law. 
Both equations can be regarded as elliptic type differential equations for the magnetic flux $\Psi(R,z)$.
Thanks to decomposing the pulsar equation into two elliptic type differential equations, we have one more
unknown function $J_{\rm T}(R,z)$ in addition to $\Psi(R,z)$ and $I_{\rm P}(\Psi)$.
Therefore, we can regard $J_{\rm T}(R,z)$ as an adjustable function rather than $I_{\rm P}(\Psi)$.
This is the most important point of difference from
the CKF method, which employed the poloidal current $I_{\rm P}(\Psi)$ as an adjustable function. 
In this paper, we consider a numerical method for obtaining a solution satisfying Eqs.~(\ref{Ampere2}) and (\ref{FF2})
simultaneously; i.e., we consider the problem of finding a set of $(\Psi, J_{\rm T})$
which satisfies Eqs.~(\ref{Ampere2}) and (\ref{FF2}) with fixed $I_{\rm P}(\Psi)$.

There seems to be no singularity like the light cylinder in Eqs.~(\ref{Ampere2}) and (\ref{FF2}).
However, the existence of the light cylinder affects the convergence of a numerical iteration. 
In fact, some iterations did not converge if the light cylinder existed in the numerical domains. 
To obtain a smooth solution across the light cylinder, we therefore construct the following iterative method:
\begin{enumerate}
 \item Set a function for $I_{\rm P}(\Psi)$ and give a trial toroidal current
       $S_{\rm Ttrial}(R,z):=8\pi^2 RJ_{\rm Ttrial}(R,z)/c$.
       We fix the functional form of $I_{\rm P}(\Psi)$ in this procedure.
 \item By solving Ampere's law (\ref{Ampere2}) with the trial $S_{\rm Ttrial}(R,z)$ numerically, 
       we obtain a magnetic flux $\Psi(R,z)$.
       In general, the obtained $\Psi$ and the trial $S_{\rm Ttrial}$ do not satisfy the force-free condition (\ref{FF2}).
 \item  Using the obtained $\Psi$ and $S_{\rm Ttrial}$, 
	we make the following new trial toroidal current $S_{\rm Tnew}(R,z)$:
       \begin{eqnarray}
	&& S_{\rm Tnew}(R,z) \nonumber \\
	&&=\left\{\frac{1+\tanh(\eta D)}{2}S_{\rm Tin}(R,z)\right. \nonumber \\
	&&~~~~\left.+\frac{1-\tanh(\eta D)}{2}S_{\rm Tout}(R,z)\right\}\left\{1-{\rm e}^{-D^2/(2\sigma^2)}\right\}
	 \nonumber \\
	 &&~~~~+S_{\rm TLC}(R_{\rm LC},z){\rm e}^{-D^2/(2\sigma^2)}, 
	 \label{JTnew}
       \end{eqnarray}
       where
       \begin{eqnarray}
	&&S_{\rm Tin}(R,z) \nonumber \\
        &&= \frac{1}{2}\left\{(1+R^2\Omega_{\rm F}^2/c^2)
			S_{\rm Ttrial}(R,z)-2\Omega_{\rm F}^2R\partial_R\Psi/c^2\right. \nonumber \\
	&&~~\left.+16\pi^2I_{\rm P}I_{\rm P}'/c^2\right\}, \label{JTin} \\
	&&S_{\rm Tout}(R,z) \nonumber \\
        &&=(1+R^2\Omega_{\rm F}^2/c^2)^{-1}\left\{2S_{\rm Ttrial}(R,z)+2\Omega_{\rm F}^2R\partial_R\Psi/c^2\right.
	 \nonumber \\
	&&~~\left.-16\pi^2I_{\rm P}I_{\rm P}'/c^2\right\},  \label{JTout} \\
	&&S_{\rm TLC}(R_{\rm LC},z) \nonumber \\
	&&=\partial_R^2\Psi|_{R=R_{\rm LC}}
	 +R_{\rm LC}^{-1}\partial_R\Psi|_{R=R_{\rm LC}} \nonumber \\
	&&~~-\frac{1}{2}R_{\rm LC}^{-1}\Omega_{\rm F}^{-2}(16\pi^2I_{\rm P}I_{\rm P}')'
	 \partial_R\Psi|_{R=R_{\rm LC}}, \\
	\label{JTLC}
	 &&D=1-R^2\Omega_{\rm F}^2/c^2,
       \end{eqnarray}
	and $\eta$ and $\sigma$ are constants. We use $S_{\rm Tnew}$
	as the next $S_{\rm Ttrial}$.
 \item Repeat these steps until the change of $S_{\rm Ttrial}(R,z)$ is numerically within a small residual.
\end{enumerate}

In this routine, we repeat to solve Ampere's law with a trial toroidal current that is constructed
by using the force-free condition. This trial current is given to make this routine converge. We performed
this routine with some trial currents and found that $S_{\rm Tin}(S_{\rm Tout})$ makes this iteration converge
if the numerical domain is inside (outside) the light cylinder. 
Therefore, by using a hyperbolic tangent and a Gaussian function, we make a suitable $S_{\rm Tnew}$ expected to make this
iteration converge in a numerical domain including the light cylinder. 
$S_{\rm Tnew}$ almost becomes $S_{\rm Tin}$ inside the light cylinder, and then the hyperbolic 
tangent switches $S_{\rm Tnew}$ from $S_{\rm Tin}$ to $S_{\rm Tout}$ outside the light cylinder. 
At the light cylinder, thanks to the Gaussian function, $S_{\rm Tnew}$ becomes $S_{\rm TLC}$ 
which gives the correct value of $S_{\rm T}(R,z)$ at the light cylinder if the light cylinder condition (\ref{LC}) is satisfied. 
We can control the behavior of $S_{\rm T}$ by determining the constants $\sigma$ and $\eta$. 
Which values should be adopted for these parameters depends on the result of this calculation, but
we find that they do not play an essential work for the global structure of magnetosphere if the iteration converges.
The derivation of $S_{\rm Tnew}$ and the detail are written in Appendix \ref{new-current}. 
In step 4, we should determine when we stop this routine. The criterion of the convergence 
is discussed in Appendix \ref{convergency}. 
Lastly, we should notify that the light cylinder condition is not imposed in this routine. 
This is different from the CKF method and it allows us to determine a
functional form of $I_{\rm P}(\Psi)$ freely because the light cylinder condition restricts 
the functional form of the poloidal current.
This is why the CKF solution has a poloidal current sheet.
Note that there is no guarantee of satisfaction of
the light cylinder condition, even if the iteration converges.
If the light cylinder condition is not satisfied, the obtained solutions do not
express isolated pulsar magnetospheres but those including other structures, such as a corona.

\subsection{Michel monopole solution} \label{monopole}
\begin{figure*}
\begin{center}
 \includegraphics[width=63mm]{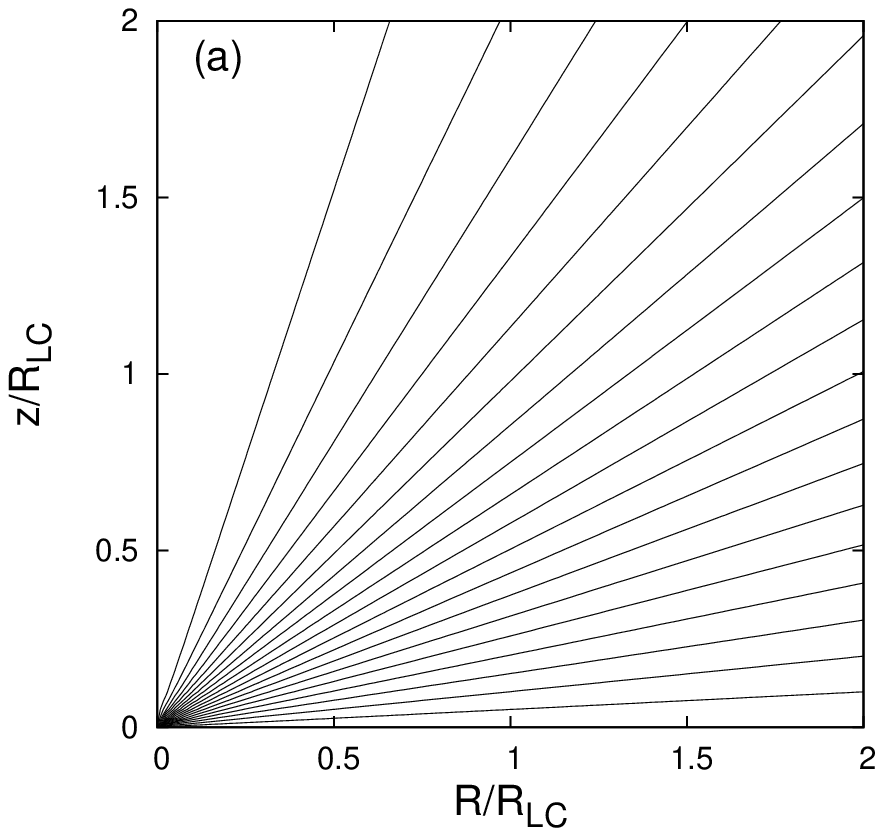}
 \includegraphics[width=63mm]{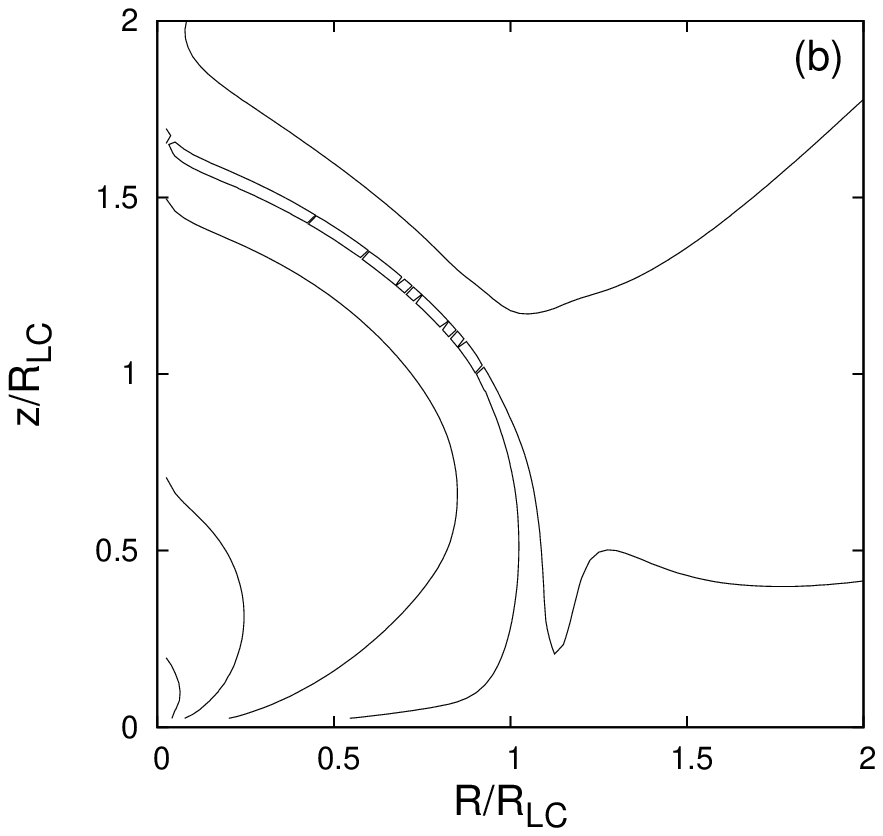}
\end{center}
 \caption{(a) Contour plot of  $\Psi$. The solid line nearest to the z-axis is the magnetic field line with $\Psi=0.05C$, 
and then the other solid lines are drawn from there by adding $\Psi=0.05C$. The magnetic field lines pass through the light
cylinder, $R=R_{\rm LC}$, smoothly. (b) Contour plot of the relative error between the numerical result and the Michel monopole solution.  
$\delta\Psi$ is defined as $\delta\Psi:=|(\Psi_{\rm N}-\Psi_{\rm M})/\Psi_{\rm M}|$ 
where $\Psi_{\rm N}$ and $\Psi_{\rm M}$ are the magnetic flux function of
our numerical result and the Michel monopole solution, respectively. We plot the contours that are 
$\delta\Psi=10^{-1}$, $10^{-2}$, $10^{-3}$, and $10^{-4}$. 
The contour line close to the origin represents $\delta\Psi=10^{-1}$ and the relative error is decreasing
with distance. Since there is the line with $\delta\Psi=10^{-3}$ near the $z$-boundary, the relative error is not monotonically
decreasing but the order is about $10^{-3}$ at most. Thus, we see that the relative error is quite small except near the
origin. Since the Michel monopole solution diverges at the origin, the relative error near the origin is 
larger than in the other region but it is about 40\% at most.}
\label{fig:mono} 
\end{figure*}

Before considering pulsar magnetospheres with a CKF-type structure, 
we check whether our routine works well.
As a test, we try to get an exact solution of the pulsar equation 
called the Michel monopole solution which is given by (see, e.g., \cite{2003PThPh.109..6190})
\begin{eqnarray}
 \Psi(R,z) &=& C\left(1-\frac{z}{\sqrt{R^2+z^2}}\right),\label{michel:psi} \\ 
 4\pi I_{\rm P}(\Psi) &=& -\Omega_{\rm F}\Psi\left(2-\frac{\Psi}{C}\right),\label{michel:Ip} \\
 J_{\rm T}(R,z) &=& 0, \label{michel:JT} 
\end{eqnarray}
where $C$ is a constant. Michel monopole solution represents a rotating magnetic monopole. 
We perform calculations in the finite domain $(0,0)<(R,z)\leq(2R_{\rm LC},2R_{\rm LC})$ divided into $80\times80$ numerical meshes. 
In addition, we impose the boundary conditions as $\Psi=0$ at the axis $R=0$, $\Psi/C=1$ at
the equator $z=0$, and  $R\partial_R\Psi+z\partial_z\Psi=0$ at the outer boundaries. 
Then, we give the poloidal current function as Eq.~(\ref{michel:Ip}).
Furthermore, we give $\eta$ and $\sigma$ in $S_{\rm Tnew}$ as
$\eta=50$ and $\sigma=0.1$ for this calculation and start our routine with 
a non-zero initial toroidal current density. Results are shown in Fig.~\ref{fig:mono}. 
We see that our result agrees with the Michel monopole solution and it is smooth over the light cylinder.

\subsection{CKF-type solutions}
We study pulsar magnetosphere models that have a CKF-type structure. Unlike the CKF method, 
we need to determine the functional form of $I_{\rm P}(\Psi)$. 
To do so, we first inspect the poloidal currents obtained in
\citet{1999ApJ...511..351C, 2003PThPh.109..6190, 2004MNRAS.349..213G, 2005PhRvL..94b1101G, 2006MNRAS.368.1055T}.
We then model the poloidal current $I_{\rm P}I_{\rm P}'$ by using a cubic function. 
We describe our numerical model and show results.

\subsubsection{Boundary Conditions}\label{bdc}
In this work, we consider a finite rectangular domain $0<R\leq R_{\rm max}$ 
and $0<z\leq z_{\rm max}$ to solve 
Eq.~(\ref{Ampere2}) and impose the boundary conditions as discussed in \citet{2006MNRAS.368.1055T}.
First, let us consider the z-axis, $R=0$. Since magnetic fields at the axis should not diverge, 
there is no magnetic flux at the axis. The appropriate boundary condition is then given by
\begin{equation}
 \Psi(0,z)=0~~{\rm for}~~0<z\leq z_{\rm max}.
\end{equation}
Then at the equator, $z=0$, in order to obtain a magnetosphere with both closed and open
magnetic field lines, we give the following boundary condition:
\begin{eqnarray}
 \partial_z\Psi(R,0) &=& 0~~{\rm for}~~0<R<R_{\rm LC}, \\
 \Psi(R,0) &=& \Psi_{\rm op}~~{\rm for}~~R_{\rm LC}\leq R \leq R_{\rm max},
\end{eqnarray}
where $\Psi_{\rm op}$ is a constant.
It is expected that the magnetic field lines are closed for $\Psi>\Psi_{\rm op}$. 
Therefore, $\Psi_{\rm op}$ gives the value of the last open field line.

Near the origin, we expect magnetospheres to be dipolar, with $\Psi$ given by
\begin{eqnarray}
 &&\Psi(R,z)=\frac{mR^2}{(R^2+z^2)^{3/2}}, \nonumber \\
&&~~~~~~~~~~~~~~{\rm for}~~(0,0)<(R,z)\leq(R_{\rm S},z_{\rm S}),
\end{eqnarray}
where $m$ is a magnetic dipole moment and $R_{\rm S}$ and $z_{\rm S}$ represent the surface of the star. 
We give this dipole's value at the surface and the interior of a neutron star.

Lastly, at the outer boundaries, we impose boundary conditions with the monopole magnetic field line 
as considered by \citet{2006MNRAS.368.1055T}:
\begin{eqnarray}
 &&R\partial_R\Psi+z\partial_z\Psi=0,~~{\rm for}~~R=R_{\rm max},~~0<z\leq z_{\rm max}\nonumber \\
 &&{\rm and}~~0<R\leq R_{\rm max},~~z=z_{\rm max}. \label{obdc}
\end{eqnarray}
For the CKF solution, other boundary conditions are imposed at the outer boundaries 
\citep{2003PThPh.109..6190}; however, their results are quite similar to the CKF solution obtained in \citet{2006MNRAS.368.1055T} 
in a region over the light cylinder. 
It is expected that physical nature of possible solutions are not 
sensitive to outer boundary conditions, at least near the neutron star. 

\subsubsection{Model of the Poloidal Current $I_{\rm P}(\Psi)$}\label{model_I}
In our method, we fix the functional form of $I_{\rm P}(\Psi)$.
In the previous works for the CKF solution, $I_{\rm P}I_{\rm P}'$ 
is used as the adjustable function rather than $I_{\rm P}$ 
and the obtained $I_{\rm P}I_{\rm P}'$ shows almost the same behavior.
By inspecting their $I_{\rm P}I_{\rm P}'$, we modeled their $I_{\rm P} I_{\rm P}'$ by the following cubic equation:
\begin{equation}
 16\pi^2I_{\rm P}I_{\rm P}'=16\pi^2 A^2\Psi(\Psi-\Psi_{\rm ret})(\Psi-\Psi_{\rm op}), 
 \label{IdI}
\end{equation}
for $0\leq\Psi<\Psi_{\rm op}$ where $A$ and $\Psi_{\rm ret}(\leq\Psi_{\rm op})$ are constants. 
Moreover, for $\Psi\geq\Psi_{\rm op}$, we assume that $I_{\rm P}I_{\rm P}'=0$ because
there is no poloidal current in the closed field zone.
Since $I_{\rm P}I_{\rm P}'=(I_{\rm P}^2)'/2$, we can easily integrate Eq.~(\ref{IdI}) 
for $0\leq \Psi \leq \Psi_{\rm op}$ and obtain
\begin{eqnarray}
 &&4\pi I_{\rm P}(\Psi) \nonumber \\
 &&=\pm 4\pi A\Psi\sqrt{\frac{1}{2}\Psi^2-\frac{2(\Psi_{\rm ret}+\Psi_{\rm op})}{3}\Psi
    +\Psi_{\rm ret}\Psi_{\rm op}}, \label{I}
\end{eqnarray}
for $0\leq\Psi<\Psi_{\rm op}$, where we imposed the boundary condition as $I_{\rm P}(0)=0$ 
so as to make the poloidal current density regular at the axis. 

For $\Psi\geq\Psi_{\rm op}$, which should be the closed field zone, we impose $I_{\rm P}(\Psi)=0$.
In this model, there are three parameters, $A$, $\Psi_{\rm ret}$, and $\Psi_{\rm op}$.
$\Psi_{\rm ret}$ should satisfy $\Psi_{\rm op}/2\leq\Psi_{\rm ret}\leq\Psi_{\rm op}$, 
so as not to make Eq.~(\ref{I}) imaginary for $0\leq\Psi\leq\Psi_{\rm op}$.

$\Psi_{\rm op}$ represents the last open field line.
Moreover, assuming that the pulsed emission of the pulsar originates from the polar cap, 
this parameter determines the emission area in the polar region near the neutron star's surface.
Let $\theta_{\rm p}$ be the polar angle of the last open field line on the star's surface.
Since we put the dipole configuration as the boundary condition on the star's surface, 
the value of the magnetic flux at $\theta_{\rm p}$ is given by $m \sin^2\theta_{\rm p}/R_{\rm S}$ and 
it corresponds to $\Psi_{\rm op}$. Thus we have
\begin{equation}
 \Psi_{\rm op}=\frac{m \sin^2\theta_{\rm p}}{R_{\rm S}}.
 \label{psi_op}
\end{equation}
When a rotating dipole is considered, $\Psi_{\rm op}$ is taken as  $\Psi_{\rm op}=m/R_{\rm LC}=:\Psi_{0}$,
which corresponds to the value of the magnetic flux of the dipole on the light cylinder at the equator. 
By using $\Psi_0$, we can rewrite Eq.~(\ref{psi_op}) as 
\begin{equation}
 \sin^2\theta_{\rm p}=\frac{\Psi_{\rm op}R_{\rm S}}{\Psi_0 R_{\rm LC}}.
\end{equation}
This equation gives the polar angle $\theta_{\rm p}$.
$\Psi_{\rm op}/\Psi_{0}$ would be of the order unity because $\Psi_0$ represents 
the typical value of the last field line for a CKF-type structure.
Considering a millisecond pulsar, $R_{\rm S}/R_{\rm LC}$ is about $10^{-2}$ since $R_{\rm S} \sim 10$ km and $R_{\rm LC}\sim 10^3$ km.
Thus the typical value of $\theta_{\rm p}$ is about $10^{-1}$ rad which is about 5.7 degrees.

Let us discuss the meanings of the other parameters.
From Eq.~(\ref{B}), the toroidal component of the magnetic field, $B_{\rm T}$, is given by
\begin{equation}
 B_{\rm T}=\frac{2I_{\rm P}(\Psi)}{R}.
\end{equation}
Therefore roughly speaking, $I_{\rm P}$ expresses the toroidal magnetic field. 
Thus, the parameter $A$ determines the strength of the toroidal magnetic field; 
the choice of sign determines the direction of the toroidal magnetic field. 
Since $I_{\rm P}$ appears in the form of $I_{\rm P}I_{\rm P}'$ in the pulsar equation, 
the choice of its sign is not essential. We  therefore choose minus, as other authors did.

Then, $\Psi_{\rm ret}$ represents the existence of return currents.
From Eq.~(\ref{IdI}), we see that $I_{\rm P}I_{\rm P}'(\Psi_{\rm ret})=0$,
and, from Eq.~(\ref{I}), $I_{\rm P}(\Psi_{\rm ret})\neq 0$ in general.
Thus $I_{\rm P}'$ becomes zero at $\Psi=\Psi_{\rm ret}$, 
that is, the sign of $I_{\rm P}'$ changes at $\Psi=\Psi_{\rm ret}$.
From Eq.~({\ref{Jp}}), the poloidal current density $\vec{J}_{\rm P}$ is given by
\begin{equation}
 \vec{J}_{\rm P}=I_{\rm P}'\vec{B}_{\rm P}.
 \label{Jp2}
\end{equation}
Thus, since $\vec{J}_{\rm P}$ is proportional to $I_{\rm P}'$,
the return currents flow for $\Psi_{\rm ret}<\Psi<\Psi_{\rm op}$.

Thanks to fixing the functional form of the poloidal current, we can perform calculations with 
different types of the poloidal current with fixed boundary conditions.
In addition, we can control the existence of a current sheet.
Choosing $\Psi_{\rm ret}$ as $\Psi_{\rm ret}=\Psi_{\rm op}/2$, 
it is easy to check that $I_{\rm P}(\Psi_{\rm op})=0$ in this case.
Moreover, tuning $\Psi_{\rm ret}=\Psi_{\rm op}$, we can consider the case that there exists 
only a current sheet, that is, there is no return current except the current sheet.

\subsubsection{How to give the parameters}
We model the poloidal current function $I_{\rm P}(\Psi)$ by using 
a cubic function and it contains three parameters $A$, $\Psi_{\rm ret}$, and $\Psi_{\rm op}$.
Although we try various sets of these parameters, the iteration does not always converge, 
which is expected by the regularity condition of the pulsar equation at the light cylinder.
Therefore, we should give suitable parameters for calculations.
Here, we discuss how we determine the parameters $A$, $\Psi_{\rm ret}$, and $\Psi_{\rm op}$.

Let us introduce the ratio $r$ as
\begin{equation}
 r:=\frac{\Psi_{\rm ret}}{\Psi_{\rm op}}.
 \label{ratio}
\end{equation}
Interestingly, in the previous works (Contopoulos et al. 1999; \cite{2003PThPh.109..6190,2006MNRAS.368.1055T}), 
the ratio is given by almost the same value that is about 0.8, though the values of $\Psi_{\rm op}$ differ from each
other\footnote{The ratio $r$ is about 0.79 for $\Psi_{\rm op}=1.36\Psi_0$ in \citet{1999ApJ...511..351C}, 
0.82 for $\Psi_{\rm op}=1.66\Psi_0$ in \citet{2003PThPh.109..6190}, 
and 0.80 for $\Psi_{\rm op}=1.23\Psi_0$ in \citet{2006MNRAS.368.1055T}.}.
It is expected that the ratio $r$ is an essential parameter for the CKF solution.
Therefore, we give the ratio $r$ rather than $\Psi_{\rm ret}$.
Since  $\Psi_{\rm ret}$ should be in the range $\Psi_{\rm op}/2\leq\Psi_{\rm ret}\leq\Psi_{\rm op}$,
the range of the ratio $r$ is $0.5\leq r\leq 1$.

\begin{figure*}
\begin{center}
 \includegraphics[width=63mm]{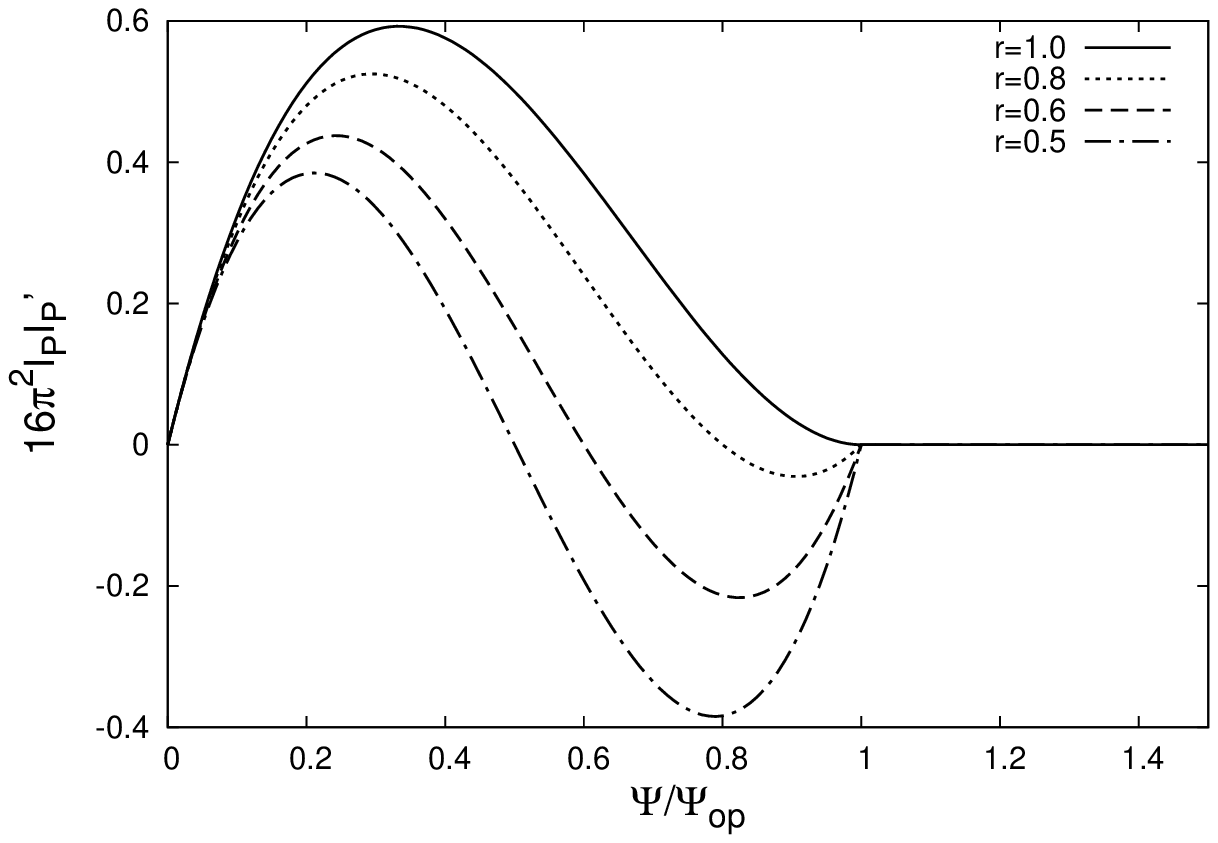}
 \includegraphics[width=63mm]{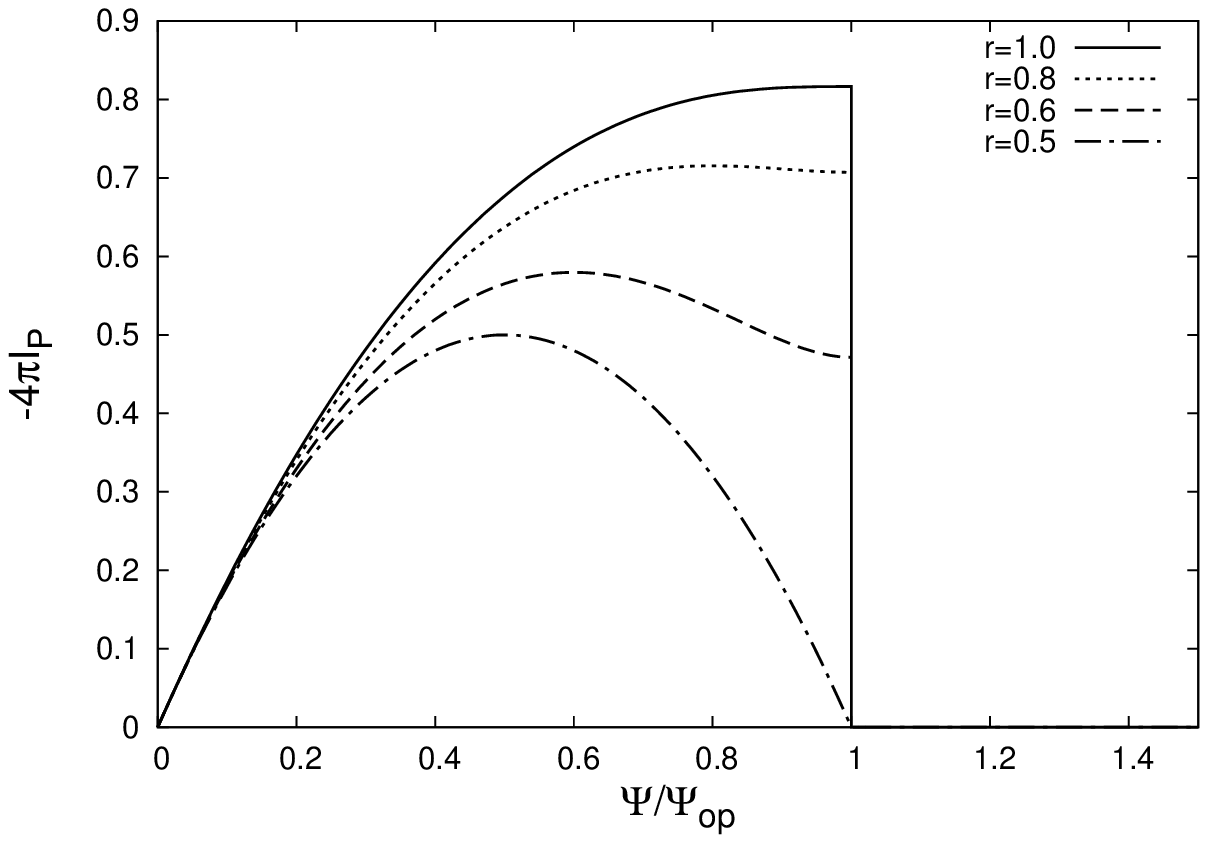}
\end{center}
 \caption{Poloidal current distribution of our model. The left panel and the right panel show 
$16\pi^2I_{\rm P}I_{\rm P}'$ and $-4\pi I_{\rm P}$ 
for the ratio $r=0.5$, 0.6, 0.8, and 1.0,  respectively.
The case with $r=0.5$ represents the model without current sheet.
The case with $r=0.8$ is the model like the CKF solution. The case 
with $r=1.0$ is the model without a return current other than a current sheet.}
\label{fig:ipol} 
\end{figure*}

Next, let us consider the parameter $A$. We investigate the poloidal current function of the CKF
solution and find that it is quite similar to that in the Michel monopole model, near $\Psi=0$. 
The Michel monopole solution gives $I_{\rm p}'(0)^2=\Omega_{\rm F}^2/4\pi^2$, 
and therefore we adjust our model to the Michel monopole solution near $\Psi=0$ by choosing $A$ as
\begin{equation}
  A^2=\frac{\Omega_{\rm F}^2}{4\pi^2\Psi_{\rm op}^2r}.
 \label{A}
\end{equation}
Finally, using given $\Psi_{\rm op}$ and $r$,
$\Psi_{\rm ret}$ and $A$ are automatically given by Eqs.~(\ref{ratio}) and (\ref{A}).
Examples of the poloidal current function are shown in Fig.~\ref{fig:ipol}.
We fix the ratio $r$ in our numerical routine, and $\Psi_{\rm op}$ is determined 
not only to make the iteration converge but also to avoid unphysical solutions. 
For example, some solutions have closed field lines beyond the light cylinder, which would be unphysical. 
Our method will not always give physically reasonable solutions even if the iteration converges.
We determine $\Psi_{\rm op}$ so as to avoid any such unphysical solutions.

\subsubsection{Results}
We are interested in constructing a pulsar magnetosphere model without 
a current sheet appeared in the CKF solution.
The existence of the current sheet and the amount
of the return current can be controlled by the ratio $r$ in $I_{\rm P}(\Psi)$ in our model. 
We perform calculations for various $r$ in the finite 
domain $(0,0)<(R,z)\leq(2R_{\rm LC},2R_{\rm LC})$ divided into $80\times80$ numerical meshes.
For the calculations, we give the value of $\eta$ and $\sigma$ in $S_{\rm Tnew}$ as
$\eta=50$ and $\sigma=0.1$. In addition, we assume that the star is
represented by the region of $(0,0)<(R,z)\leq(0.05R_{\rm LC},0.05R_{\rm LC})$,
which means that the pulsar period is about 4\,msec, assuming the star radius to be 10\,km.
We perform our numerical routine with the convergence criterion discussed 
in Appendix~\ref{convergency}.
Results are shown in Fig.~\ref{fig:psi_cont}, from which
we see that the magnetic field configuration near the neutron star is  
similar to the CKF solution for all ratios. 
However, their global structures are different. In the case with $r=0.8$, i.e., the model like the CKF solution, its structure is similar to
the CKF solution in the whole domain. On the other hand, in the case without a current sheet, i.e.,
$r=0.5$, the magnetic field becomes jet-like near the z-outer boundary inside the light cylinder. This
is similar to the result in \citet{2006ApJ...652.1494L} in which free boundary conditions are
imposed at the outer boundaries, which differ from ours. Their solution shows a disk wind near the
equatorial plane in addition to a jet flow along the axis, but our result has a quasi-spherical structure
outside the light cylinder rather than a disk wind.
\begin{figure*}
\begin{center}
 \includegraphics[width=63mm]{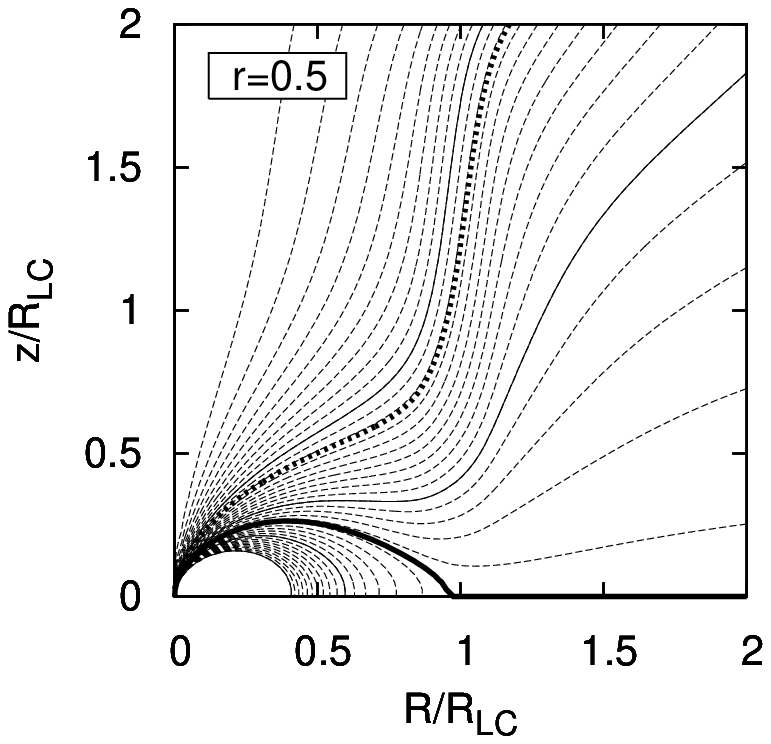}
 \includegraphics[width=63mm]{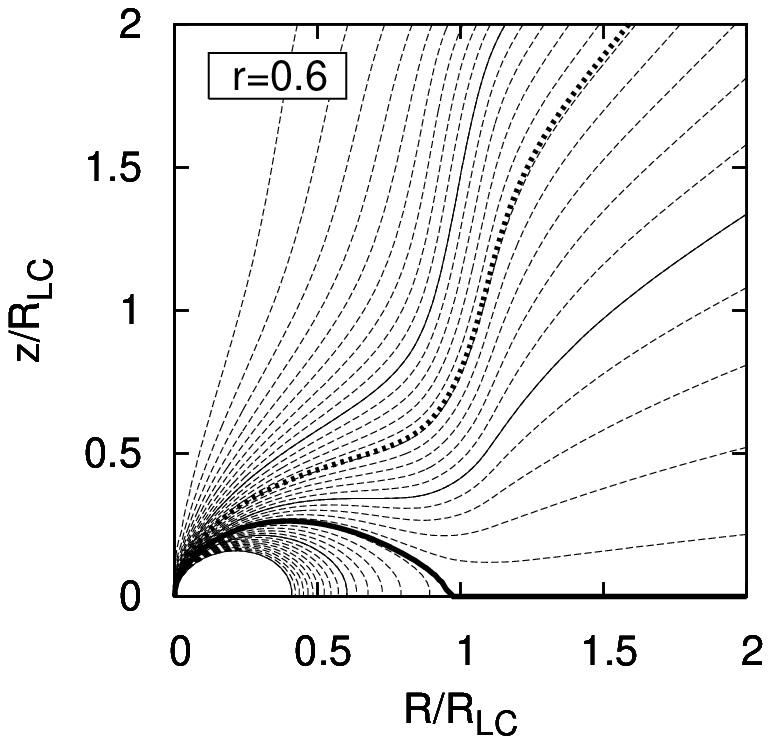}\\
 \includegraphics[width=63mm]{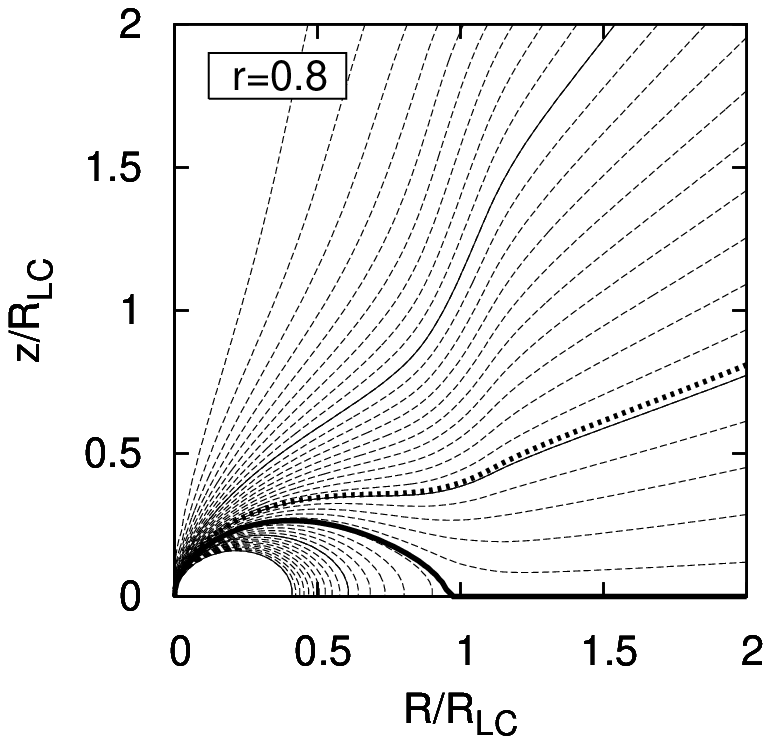}
 \includegraphics[width=63mm]{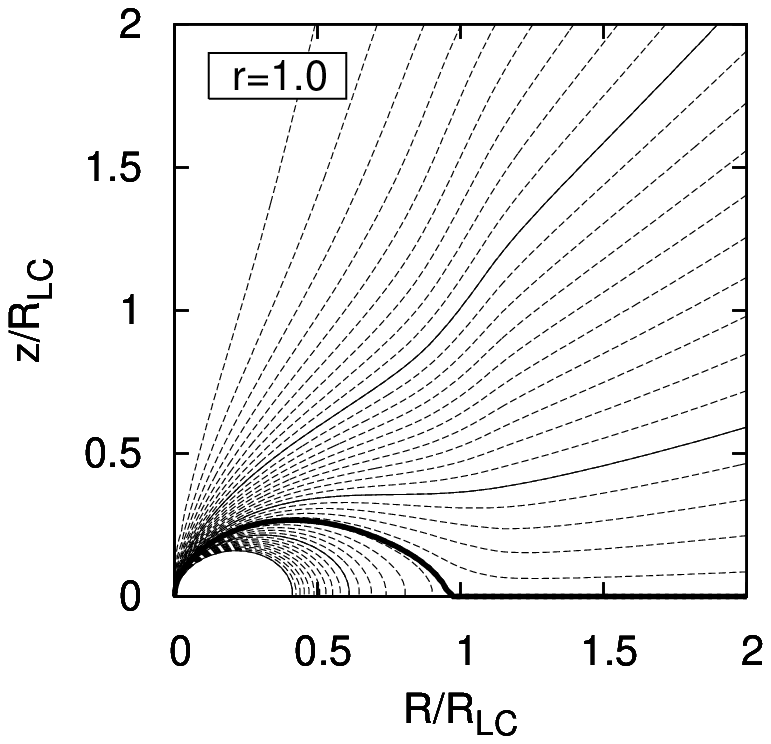}
\end{center}
 \caption{Contour plots of the magnetic flux $\Psi$. 
We show the poloidal magnetic field structures for $r = 0.5$, 0.6, 0.8, and 1.0. 
The thin dashed and solid lines in each panel represent the magnetic field lines. The thin dashed line closest
to the z-axis is the magnetic field line with $\Psi=0.05\Psi_0$, and then the values of the thin dashed lines and the
thin solid lines are increasing from there with $0.05\Psi_0$ and $0.5\Psi_0$, respectively. The thick solid line in each panel
represents the last open field line: $\Psi_{\rm op}=1.225\Psi_0$  0 for $r = 0.5$,  $\Psi_{\rm op}=1.234\Psi_0$ for $r = 0.6$,  
$\Psi_{\rm op}=1.236\Psi_0$ for $r = 0.8$, and $\Psi_{\rm op} = 1.234\Psi_0$ for $r = 1.0$. 
The return current flows between the thick dotted line and the last open field line in each
panel except the case with $r = 1.0$. In the case with $r = 1.0$, there is no return 
current except the current sheet.}
\label{fig:psi_cont} 
\end{figure*}
\begin{figure*}
 \begin{center}
  \includegraphics[width=55mm]{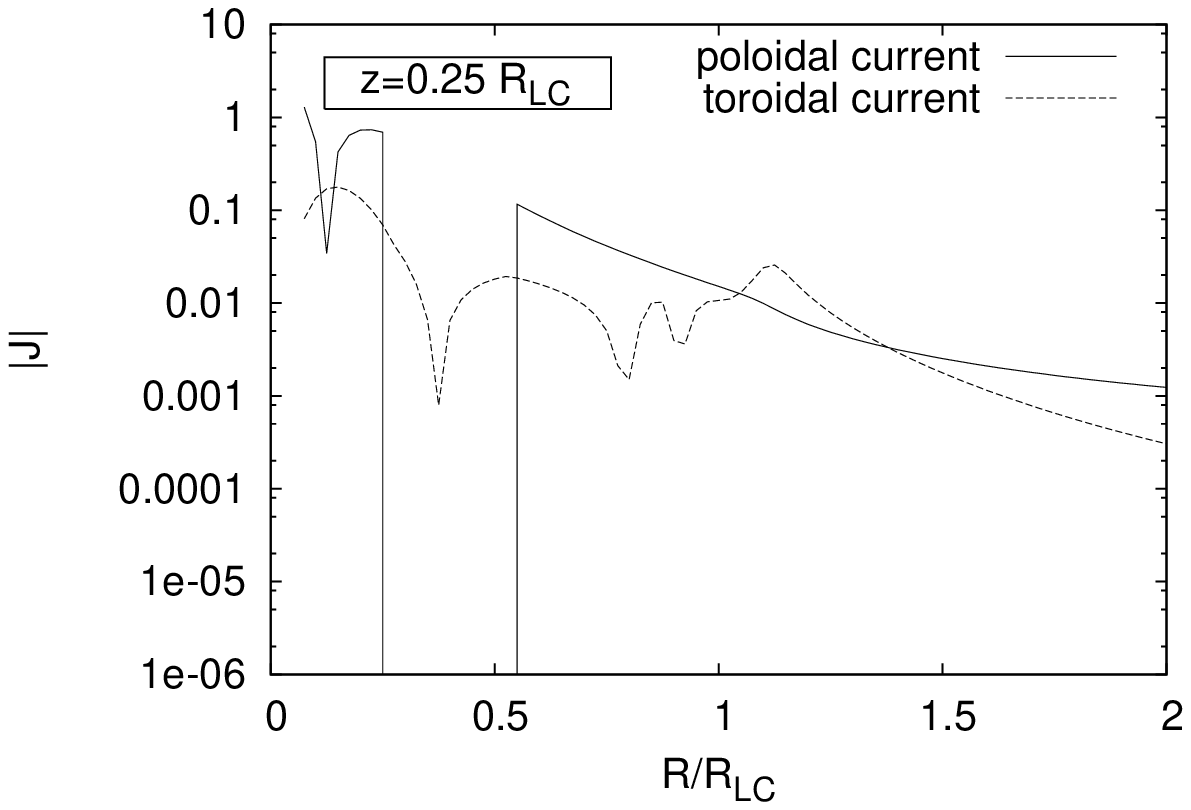}
  \includegraphics[width=55mm]{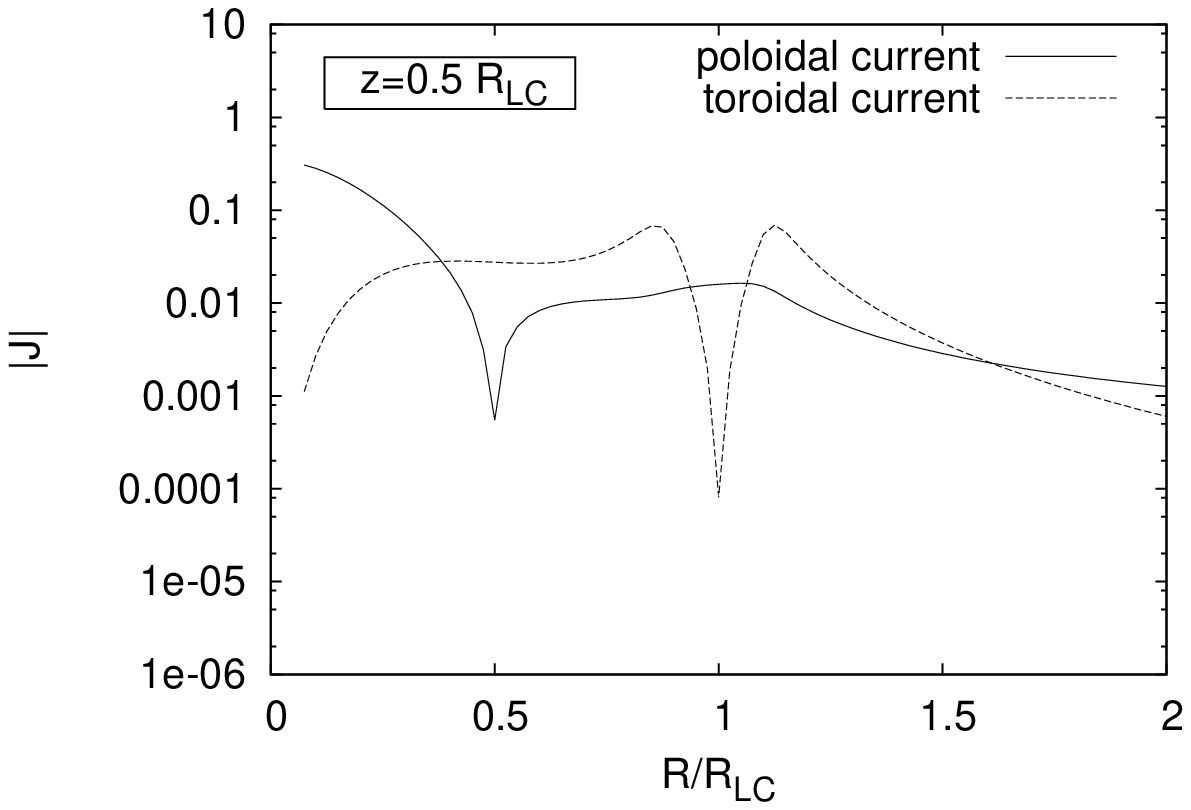}
  \includegraphics[width=55mm]{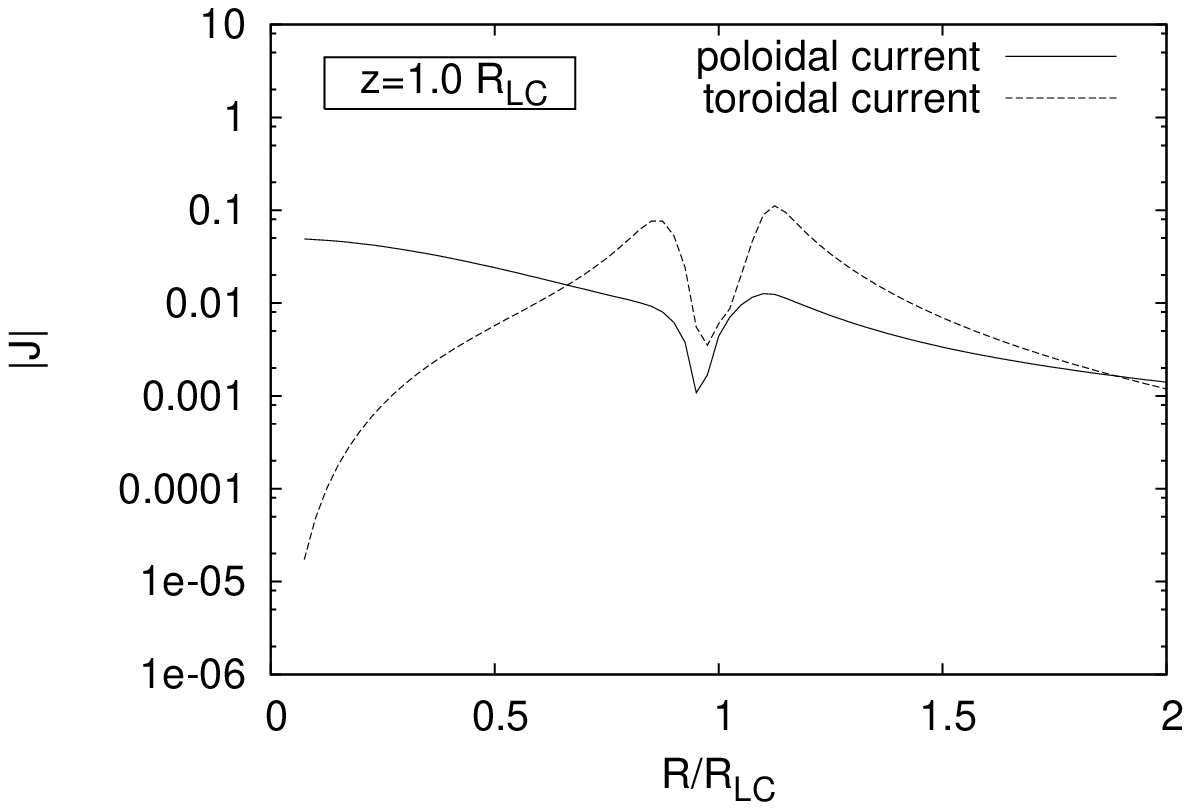}\\
  \includegraphics[width=55mm]{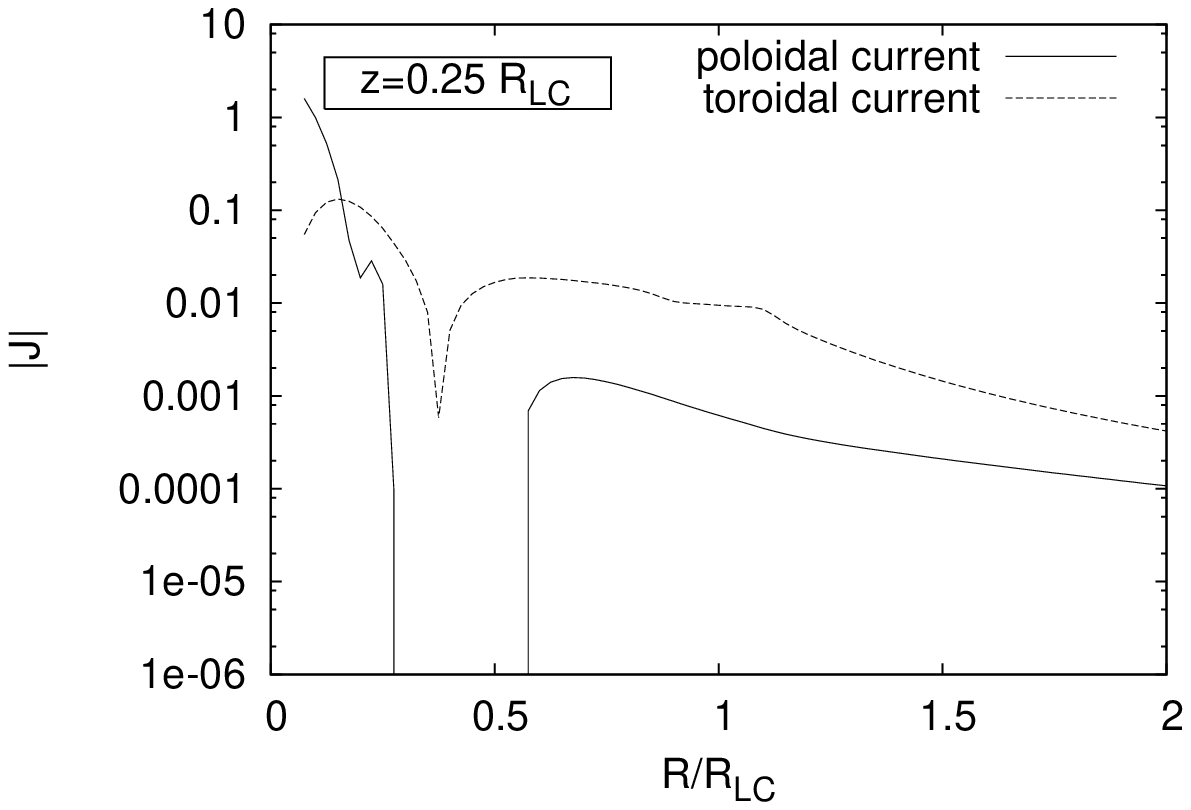}
  \includegraphics[width=55mm]{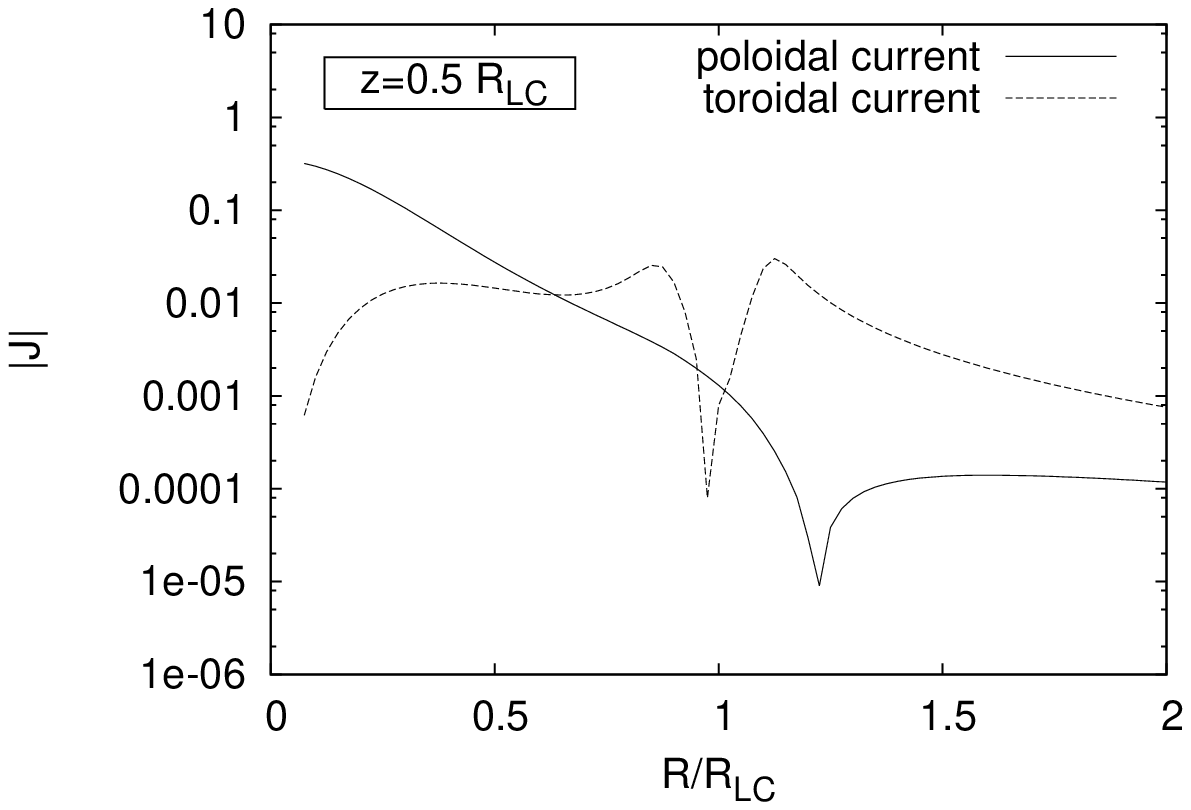}
  \includegraphics[width=55mm]{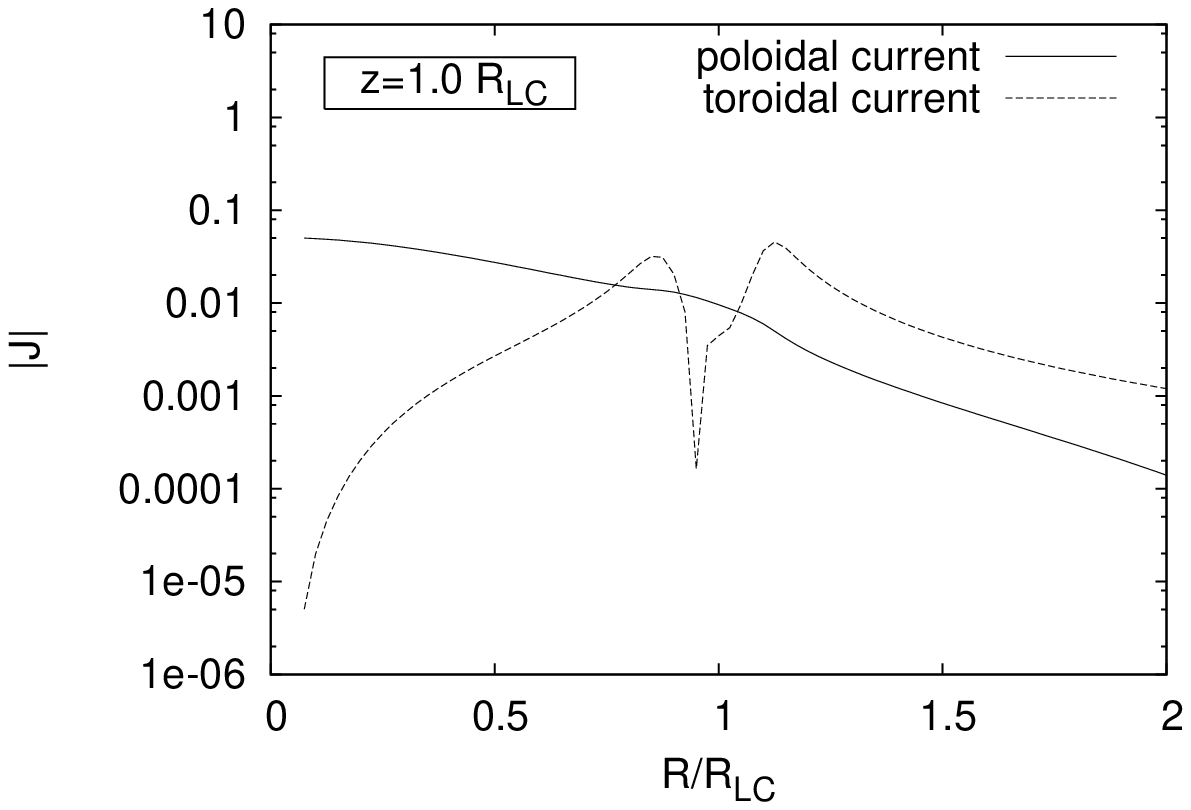}
 \end{center}
 \caption{The R-dependence of the magnitudes of the poloidal and toroidal current density. 
The top panels represent those at $z=0.25R_{\rm LC}$, $z=0.5R_{\rm LC}$, 
 and $z=1.0R_{\rm LC}$ in the case without a current sheet, that is, $r=0.5$ . 
The bottom panels show the case with $r = 0.8$ which is the case like the CKF solution and with a current sheet. Because we show
the magnitude of the electric current, there are some sharp points in the figures, which indicate the change of sign of
the electric current. The rapid decrease of the poloidal currents in the left panels indicates the existence of the
dead zone.
}
\label{fig:jt} 
\end{figure*}

One might think that a large toroidal current density appears near 
the light cylinder in the case without a current sheet, but it seems to
have no clear relation between the jet-like structure and the toroidal current.
We compare the magnitudes of the poloidal and the toroidal current density 
for $r=0.5$ and $0.8$. Results are shown in Fig.~\ref{fig:jt}.
We see that there is no large difference between the 
poloidal and the toroidal current density in the both ratios.
Moreover, a large toroidal current density is not needed to make a jet-like structure.
This is because we know an exact solution of Ampere's law in the vacuum whose magnetic field 
is only the z-component, that is, an uniform magnetic field, which might be the origin of the 
strong bending of the magnetic field around the light cylinder.
A jet-like structure can appear without a toroidal current density
though we do not answer why the jet-like structure is formed in the case
without a current sheet.
\section{Physical properties of the results}\label{property}
In this section, we investigate physical properties of our results. First, we calculate the
energy loss rate which is determined by integrating $I_{\rm P}(\Psi)$. 
In addition, we compare the total energy of the magnetosphere among
models we obtained in this study. Secondly, we study the difference of the
global magnetic field structure between the cases with and without a current sheet. Lastly, we check if
the force-free approximation is satisfied or breaks down in our results.

\subsection{Energy loss rate of the magnetosphere}
We obtain solutions of the pulsar magnetosphere both with and without a current sheet.
Here, let us compare the energy loss rates that determines the power of a radio pulsar.
In the force-free system, the energy carries away 
in the form of the Poynting flux. The Poynting flux, $\vec{P}$, is given by
\begin{equation}
 \vec{P}=\frac{c}{4\pi}\vec{E}\times\vec{B}.
\end{equation}
Since the Poynting flux satisfies divergence-free condition, 
the total energy carried away is given by a surface integral on a closed 2-surface $\cal A$:
\begin{equation}
 W=\int_{\cal A}\vec{P}\cdot d\vec{S}.
\end{equation}
After some calculations (see, e.g., \cite{2006MNRAS.368.1055T}), it can be written in the following form,
\begin{equation}
 W=2\int^{\Psi_{\rm op}}_{0}\frac{1}{2\pi c}\Omega_{\rm F}I_{\rm P}(\Psi)d\Psi,
 \label{energy_loss}
\end{equation}
where the factor of 2 came from the fact that there are northern and southern hemispheres
\footnote{Our formula of energy loss rate is slightly different from the formula 
in \citet{2006MNRAS.368.1055T}.
It comes from the difference of the definitions of $I_{\rm P}$ and $\Psi$. It is just notation and
there is no essential difference.}.
Finally, the energy loss rate is determined by integrating $I_{\rm P}(\Psi)$.
From Fig.~\ref{fig:ipol}, it is easy to find that
the area bounded by $I_{\rm P}(\Psi)$ is the smallest for $r=0.5$, the case without a current sheet, and then it
monotonically increase with increasing ratio $r$ up to unity if $\Psi_{\rm op}$ for all ratios are the same. For 
our numerical results, the values of $\Psi_{\rm op}$ are in fact almost the same.
Thus, the energy loss rate in the case without a current sheet would be the smallest
in our model. In fact, we calculate the energy loss rates for our results and show 
them in Table~\ref{tab:energy_loss} where $W_0$ is equal to $-m^2c/4\pi^2R_{\rm LC}^4$ and, 
for the usual dipole radiation, the maximum power represents $(2/3)W_0$.
\begin{table}
\label{tab:energy_loss}
\caption{The energy loss rate for our results.}
\begin{center}
 \begin{tabular}{ccccc}
 \hline 
 Ratio & $r=0.5$ & $r=0.6$ & $r=0.8$ & $r=1.0$ \\
 $W/W_0$ & 0.5002& 0.6797 & 0.8228 & 0.8909\\
 \hline
 \end{tabular}
\end{center}
\end{table}
Considering pulsar physics, the energy loss rate affects
the characteristic age of a pulsar.
Our result implies the characteristic age of a pulsar
depends on the existence of a current sheet, but the difference is about the factor of 2 at most.

\begin{figure}
 \begin{center}
  \includegraphics[width=84mm]{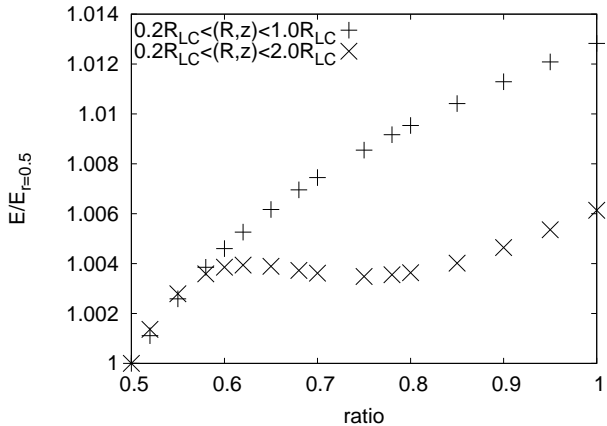}  
 \end{center}
 \caption{Comparison of  the total energies. We normalize each total energy by the case with $r=0.5$.
We estimate the total energies in the domains 
 $(0.2R_{\rm LC},0.2R_{\rm LC})<(R,z)<(1.0R_{\rm LC},1.0R_{\rm LC})$ and 
 $(0.2R_{\rm LC},0.2R_{\rm LC})<(R,z)<(2.0R_{\rm LC},2.0R_{\rm LC})$. 
The difference of the total energy is about 1.2\% at most
and becomes smaller as we take the wide region.}
 \label{fig:energy}
\end{figure}

In addition to the energy loss rate, we can also 
calculate the total energy of electromagnetic field which 
is given by the volume integration as
\begin{equation}
 {\cal E}=\int\frac{1}{8\pi}\left(E^2+B^2\right)dV.
\end{equation}
Results are shown in Fig.~\ref{fig:energy}.
Although the case without a current sheet has the least energy, the difference from other models is quite small.
This is because the magnetic field structures at the inner and 
outer boundary are irrespective of the ratio parameter $r$ in $I_{\rm P}(\Psi)$. 
The integrand is dominated by these boundary values. 
At the inner boundary, the energy density is much larger than outer part 
because we impose the boundary condition as the magnetic field structure to be dipolar, while at the
outer boundary the volume element is much larger than the inner part:
$|B|\sim 1/r^3$ near the origin where $r=\sqrt{R^2+z^2}$; 
$|B|\sim 1/r$ at a distance due to the toroidal magnetic field; and the volume element
is proportional to $r^2$.
Therefore, the integrated energies become similar for all models. 
Although the difference of the total energies is quite small, 
the energy loss rate without a current sheet is clearly smaller 
than those with a current sheet. 
Thus, the model without a current sheet seems most 
plausible in our models.

\subsection{Three dimensional magnetic field structure}
\begin{figure*}
 \begin{center}
  \includegraphics[width=41mm]{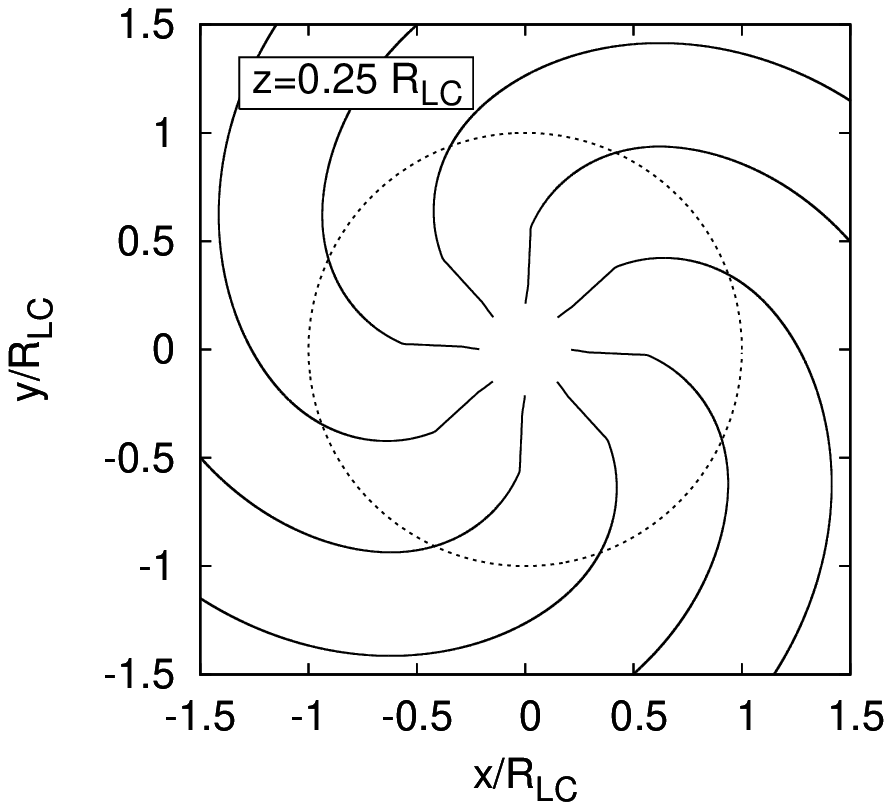}
  \includegraphics[width=41mm]{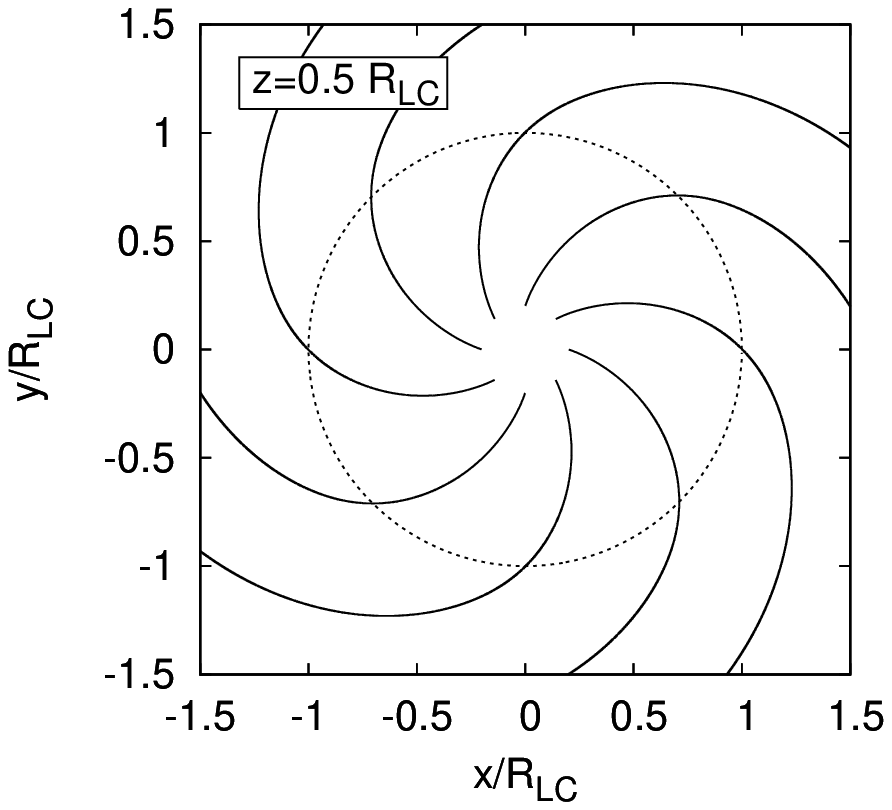}
  \includegraphics[width=41mm]{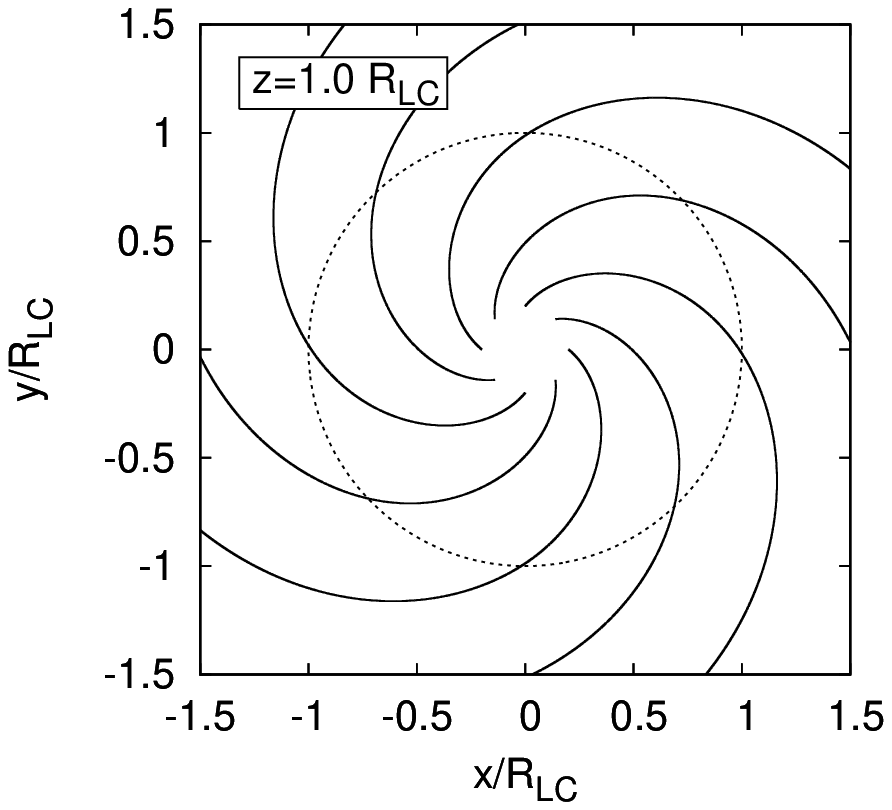}\\
  \includegraphics[width=41mm]{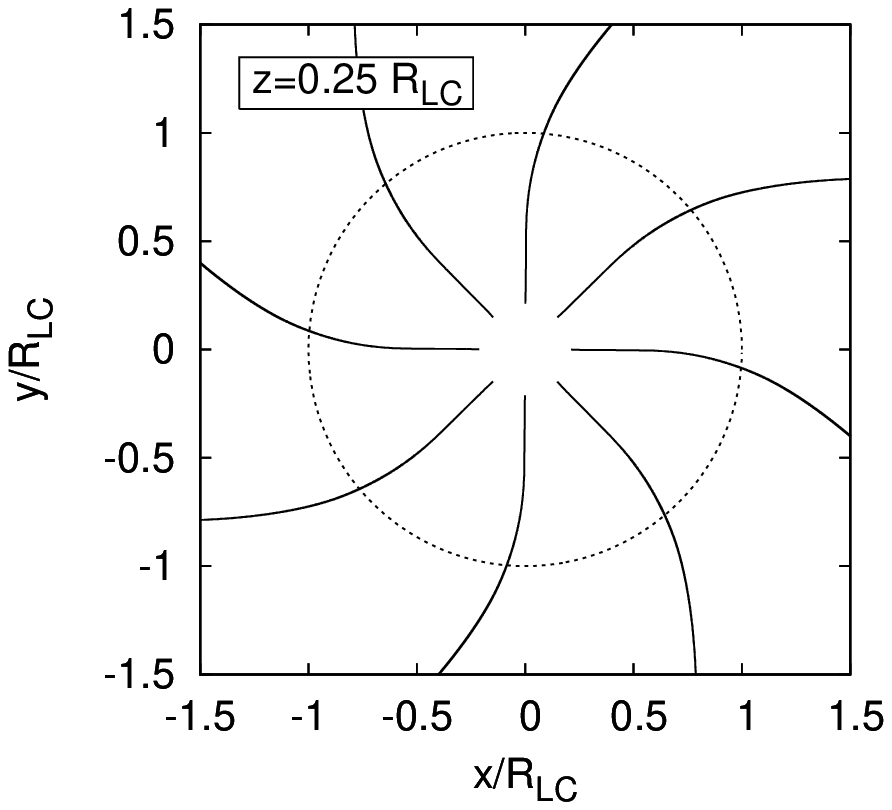}
  \includegraphics[width=41mm]{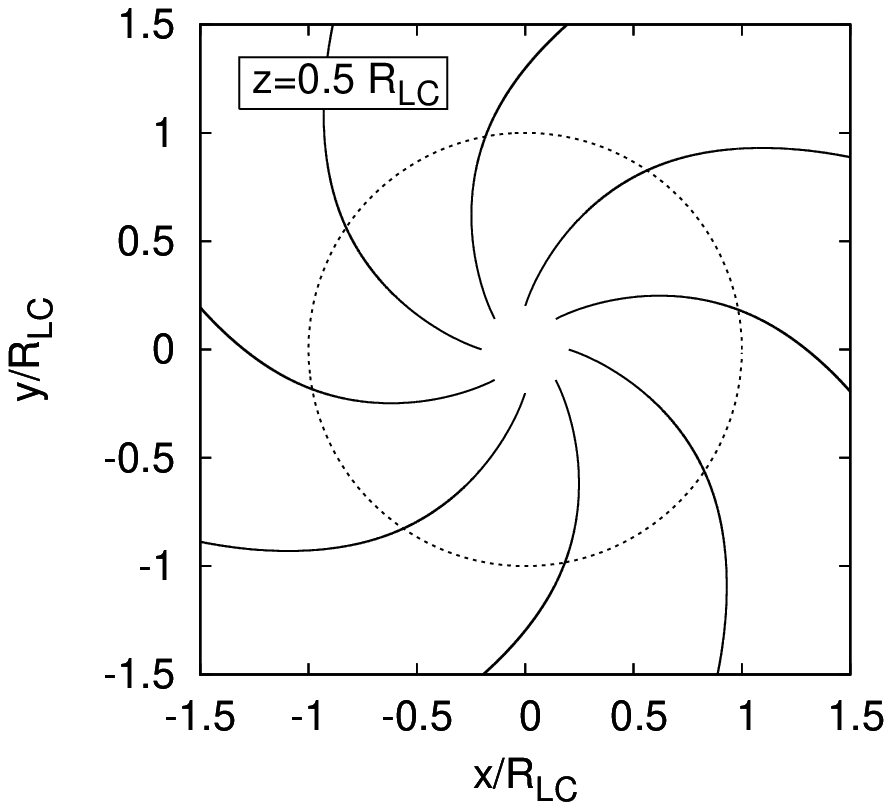}
  \includegraphics[width=41mm]{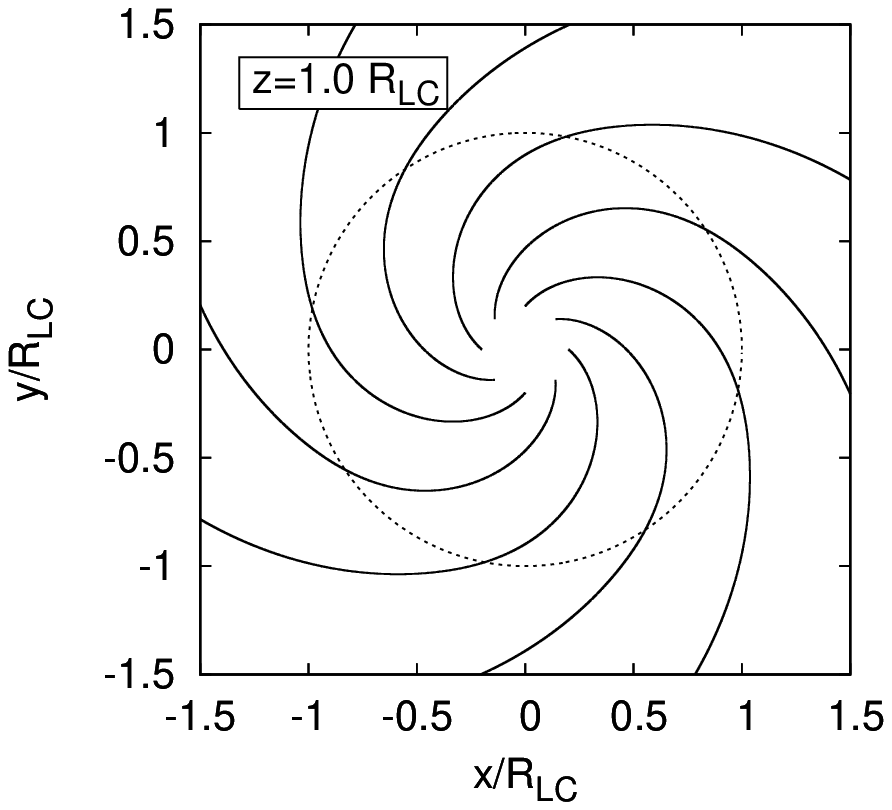}
 \end{center}
 \caption{
Magnetic field structure on an (x,y)-plane. The top panels represent the magnetic field structure in the case with $r=0.8$, 
and then the  bottom panels represent those of the case with $r=0.5$. These figures show the 
 magnetic field structure on the $(x,y)$-plane with $z=0.25R_{\rm LC}$, $z=0.5R_{\rm LC}$,
 and $z=1.0R_{\rm LC}$.
 The circle in the center of each panel indicates the light cylinder. We see that the magnetic field
 structure with/without a current sheet is quit different from each other near the equator. 
 The magnetic field lines in the case with a current sheet are more curved than those of the 
 case without a current sheet.}
\label{fig:maki2d}
\end{figure*}

\begin{figure*}
 \begin{center}
  \includegraphics[width=50mm]{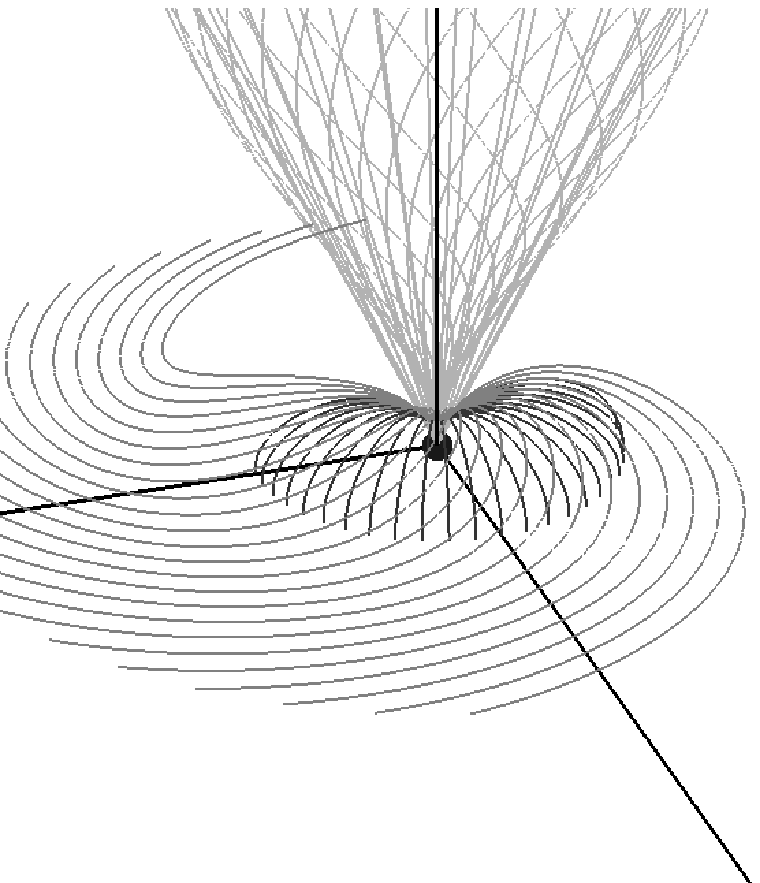}
  \includegraphics[width=50mm]{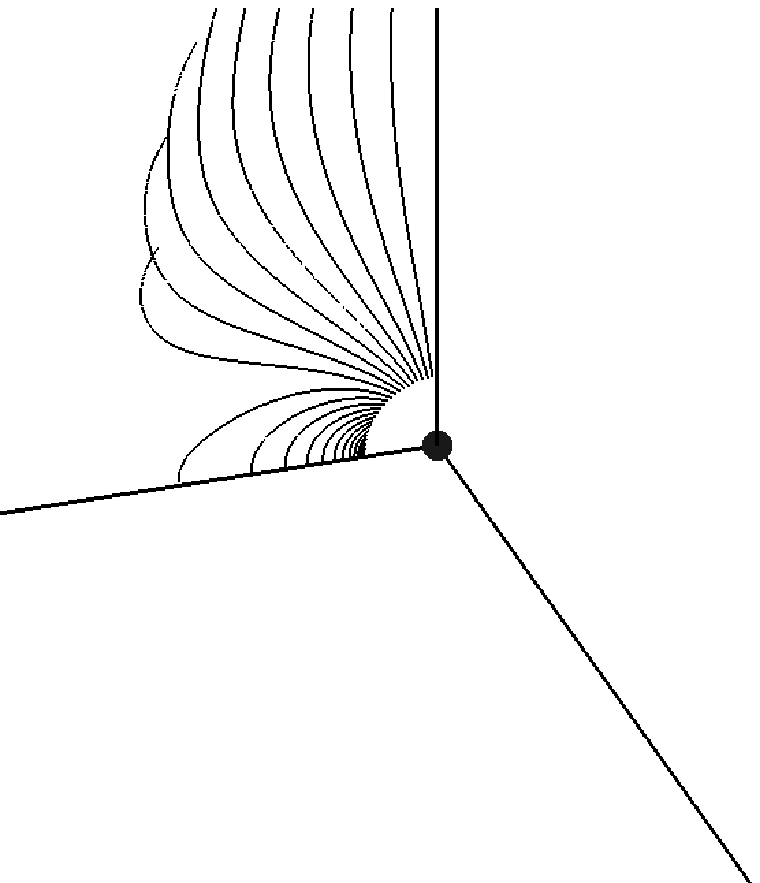}\\
  \includegraphics[width=50mm]{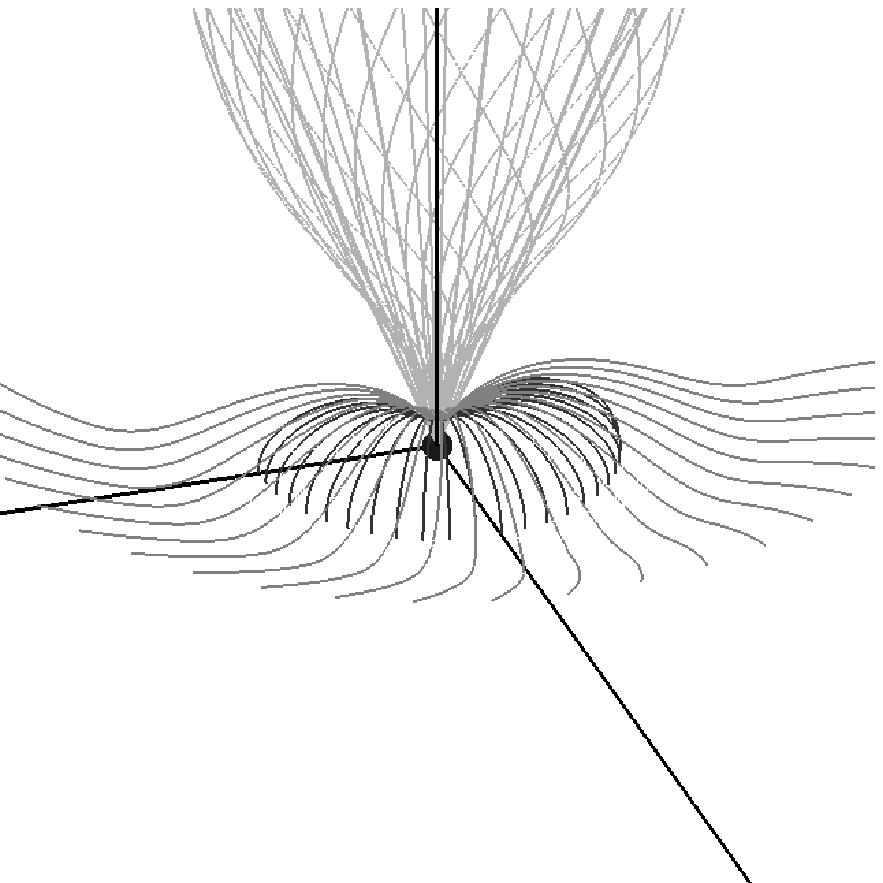}
  \includegraphics[width=50mm]{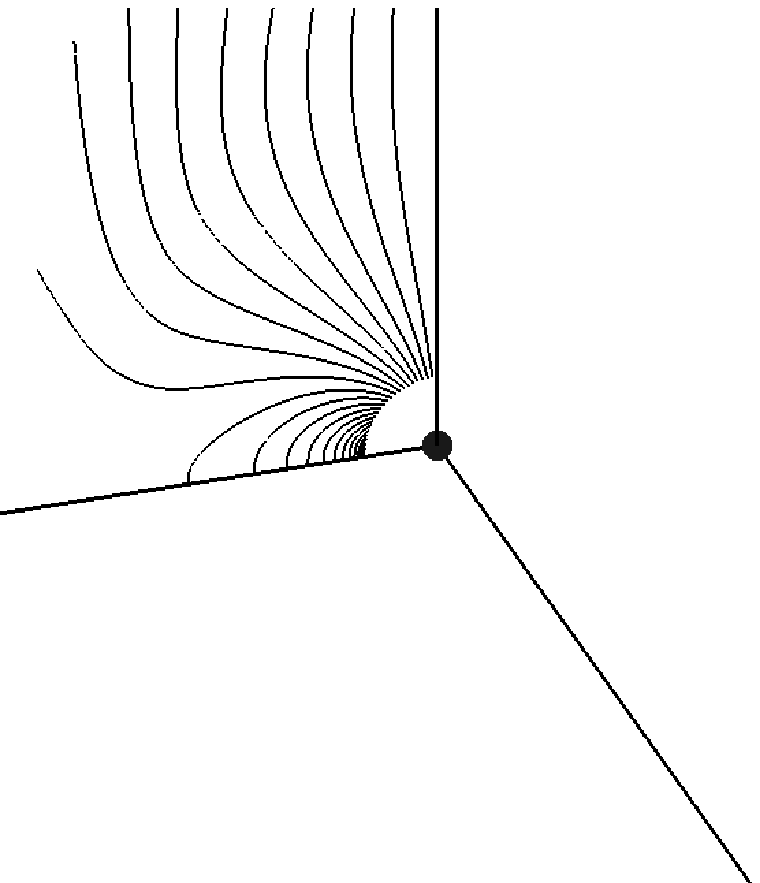}
 \end{center}
 \caption{Magnetic field structure in an $(x,y,z)$-box. 
The top panels represent  the magnetic field structure in the case with $r=0.8$,
 and then the bottom panels represent those of the case with $r=0.5$.
 Three straight lines and the black ball in the center in each panel 
 indicate the x, y, and z axes and a neutron star, respectively.
The magnetic field lines in the left panels are drawn from various points: 
$(R,\varphi,z)=(0.02R_{\rm LC},i\pi/12,0.14R_{\rm LC})$, $(0.045R_{\rm LC},i\pi/24,0.115R_{\rm LC})$, 
and $(0.06R_{\rm LC},i\pi/24,0.12R_{\rm LC})$ where $i=0,1,2,\cdots,24$.
 For the right panels, we give the initial points 
 at $(R,\varphi,z)=(0.25\sin\theta R_{\rm LC},0,0.25\cos\theta R_{\rm LC})$
 where $\theta$ is the angle from z-axis and it takes as $\theta=i\pi/48$ for $i=0,1,2,\cdots,24$.}
 \label{fig:maki3d}
\end{figure*}

The magnetic field structure is important when one considers a mechanism of
radiation from pulsar, e.g., curvature radiation.
Therefore, it is worth showing the three dimensional structure of the magnetospheres.
One magnetic field line can be drawn by integrating $\vec{B}$ which is given by Eq.~(\ref{B_ff}).
Here, we show the three dimensional magnetic field structure with two methods;
the structure on $z={\rm const}.$ plane, and the structure in a three dimensional box.
For this purpose, we change the cylindrical coordinates $(R,\varphi,z)$ to the Cartesian coordinates $(x,y,z)$ 
and integrate $\vec{B}$ on a $(x,y)$-plane or in a $(x,y,z)$-box. 
We draw the magnetic field structures in the cases with $r=0.5$ 
and 0.8 and results are shown in Figs~\ref{fig:maki2d} and \ref{fig:maki3d}.
We see that the magnetic field lines in the case with a current sheet 
are more twisted than those of the case without a current sheet.
This is consistent with the discussion of the energy loss rate because the radiation 
gets stronger for the magnetic fields with smaller curvature radius.
Because of the difference of $I_{\rm P}(\Psi)$, the three dimensional structure is 
quite different from each other. 

\subsection{Violation of the force-free approximation}
\begin{figure}
 \begin{center}
  \includegraphics[width=84mm]{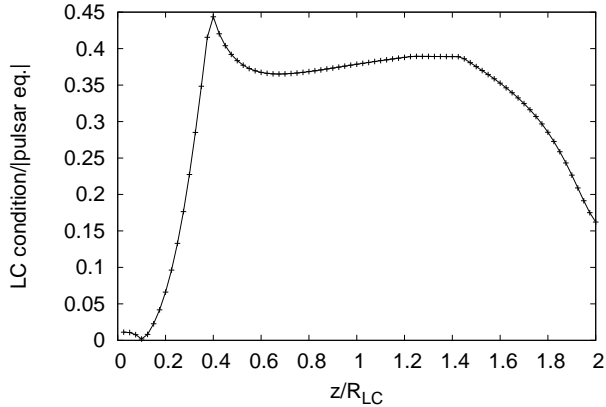} 
 \end{center}
 \caption{
Violation of the force-free condition in the case without a current sheet. We estimate the magnitude of the 
pulsar equation (\ref{eq:pulsar}) at the light cylinder, that is, the light cylinder condition (\ref{LC}). The vertical axis shows the 
magnitude along the light cylinder which is  normalized by the summation of the absolute value of each term in (\ref{eq:pulsar}). 
 The dots are the data of our results and they are connected with line.}
 \label{fig:lc_r05}
\end{figure}

Since we numerically solve the decomposed equations, it is worth to check 
whether our results satisfy the pulsar equation, in particular, at the light cylinder. 
We check it and find that our results satisfy the pulsar equation well except around the light cylinder.
The violation of the force-free condition at the light cylinder
in the case with $r=0.5$ is shown in Fig.~\ref{fig:lc_r05}, for example.
Such violation is seen around the light cylinder that is about $0.9R_{\rm LC}<R<1.1R_{LC}$
and it becomes maximum at the light cylinder.
This violation is seen in all ratios in almost the same region, 
but the violation is most suppressed in the case with  $r=1.0$.
In addition, we tried other outer boundary conditions given by \citet{2003PThPh.109..6190} and perform the same calculation. 
Although the violation of the force-free condition appeared also in this case, the degree
of the violation is improved comparing with
the boundary condition discussed in \S\ref{bdc}.
Moreover, we have assumed that the functional form of $I_{\rm P}(\Psi)$
is given by Eq.~(\ref{IdI}). It is different from the poloidal current of the CKF solution,
and therefore our solutions are different from the CKF solution in this sense.
The violation of the force-free condition in our results would be improved 
if we gave the poloidal current of the CKF solution.
To sum up, we could make the violation of the force-free condition
improved by adjusting outer boundary conditions 
and the poloidal current function though we do not know how to give 
them suitably.

We expect that the light cylinder condition would be automatically 
satisfied when our routine is finished.
In fact, we can obtain the Michel monopole solution as shown
in \S\ref{monopole}.
However, whether the light cylinder condition is satisfied
after the iteration converges is unfortunately not always guaranteed in our method.
It is because we use the light cylinder condition (\ref{LC}) to give $S_{\rm TLC}$ though
it is not satisfied during the iteration.
Concerning the astrophysical application, it is widely known that the fast-magnetosonic
point is well inside the light cylinder for most pulsars other than the Crab pulsar, whose fast-magnetosonic
point is nearly equal to or slightly further than the light cylinder. For example, the 
Lorentz factor of the fast-magnetosonic point can be written in
$\gamma_{fm} = \sqrt{\sigma}$ where
$\sigma$ represents the ratio of the Poynting energy flux to the matter energy flux
and is given by $\sigma \equiv B^2 / 4 \pi m_e n \gamma$ where
$n$ is the particle number density 
in the pulsar rest frame, $m_e$ is the electron mass, and 
$\gamma$ is the Lorentz factor of the bulk velocity \citep{1999JPhG...25R.163K}. 
By evaluating the number density by the Goldreich-Julian density 
\citep{1969ApJ...157..869G, 2009ASSL..357..421K}, 
we can estimate the fast-magnetosonic velocity at the light cylinder for various pulsars. 
In the case of the Crab pulsar, 
the ratio of Lorentz factor of fast-magnetosonic velocity and observationally estimated value 
at the light cylinder is $\sim 0.98$; 
for other pulsars whose parameters are observed or well estimated \citep{2011ApJ...741...40T}, 
the ratio is typically order of $10^{-6}$, which means the fast-magnetosonic point is at the well inside of the light cylinder
for most observed pulsars.
The force-free approximation would break at the fast-magnetosonic 
point so that the breakdown of our results at some point near the 
light cylinder is not a serious problem and can be applied to many astrophysical pulsar magnetospheres. 
In particular, in the case without a current sheet, our result has a jet-like 
structure inside the light cylinder and it gives us another possibility of CKF-type solutions.

\section{Summary and Discussion}\label{summary}

We numerically studied stationary pulsar magnetospheres in the force-free system 
both with and without a poloidal current sheet which appears in the CKF solution.
In this study, we expressed the existence of the current sheet by the ratio $r$ that 
represents the amount of the return current and constructed pulsar mangetosphere
models for various ratios.
In the case with a current sheet, in particular near $r=1.0$, 
the structures on $(R,z)$-plane are similar to the CKF solution.
Meanwhile, in the case without a current sheet, i.e. $r=0.5$, it has a jet-like structure inside the light cylinder 
and becomes quasi-spherical at a distance, which is different from the CKF solution 
but similar to the solution in \citet{2006ApJ...652.1494L}.
Then, we studied physical properties of our results
and showed that the existence of the current sheet 
affects the properties of the pulsar magnetosphere.
First, we calculated the the energy loss rates for our results
and showed that it depends on the ratio $r$.
It would affect the characteristic age of a pulsar.
Moreover, the model without a current sheet has the minimum energy 
and energy loss rate though there is no large difference for the total energies.
It appears that the model without a current sheet is plausible.
Next, we showed their three dimensional structures both with and without a current sheet. 
From Figs.~\ref{fig:maki2d} and \ref{fig:maki3d}, 
we can find that the degree of curvature of the magnetic field lines hardly depends on 
the value of $z$ in the case with a current sheet.
In the case without a current sheet, the magnetic field lines are more bent as $z$ increases.

\begin{figure*}
\begin{center}
 \includegraphics[width=63mm]{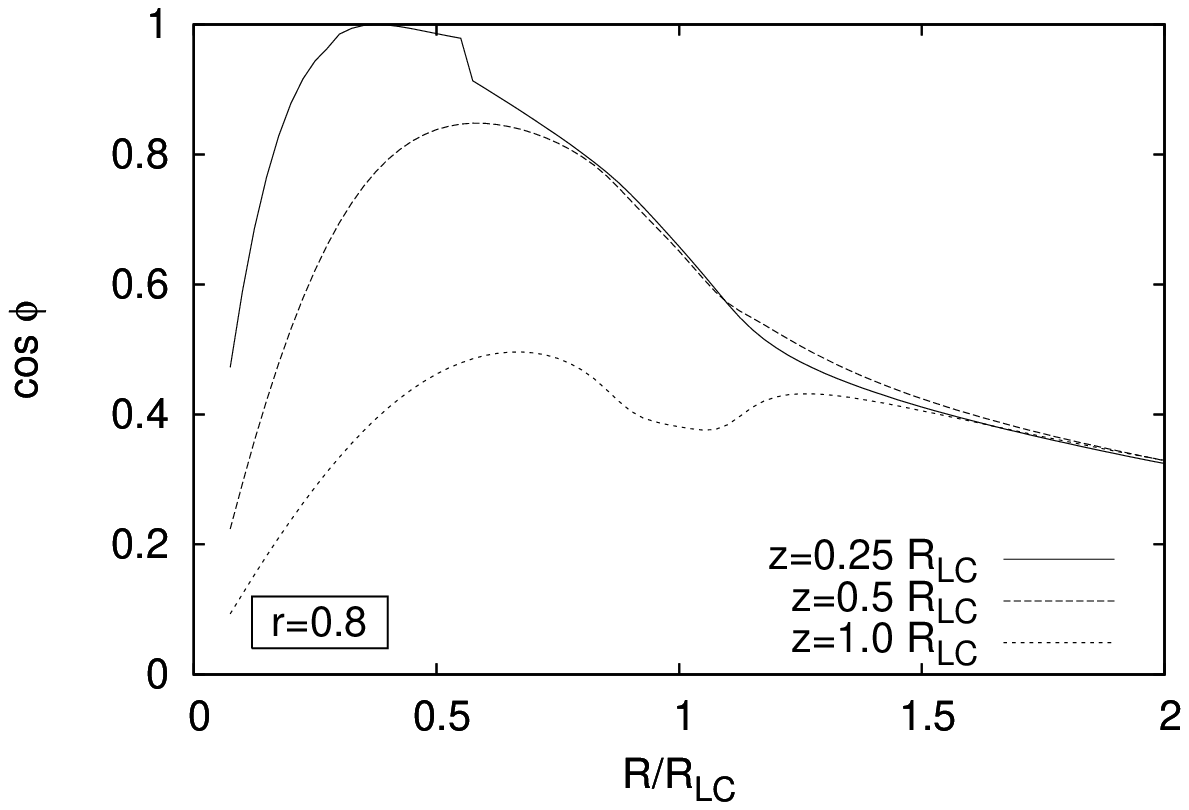}
 \includegraphics[width=63mm]{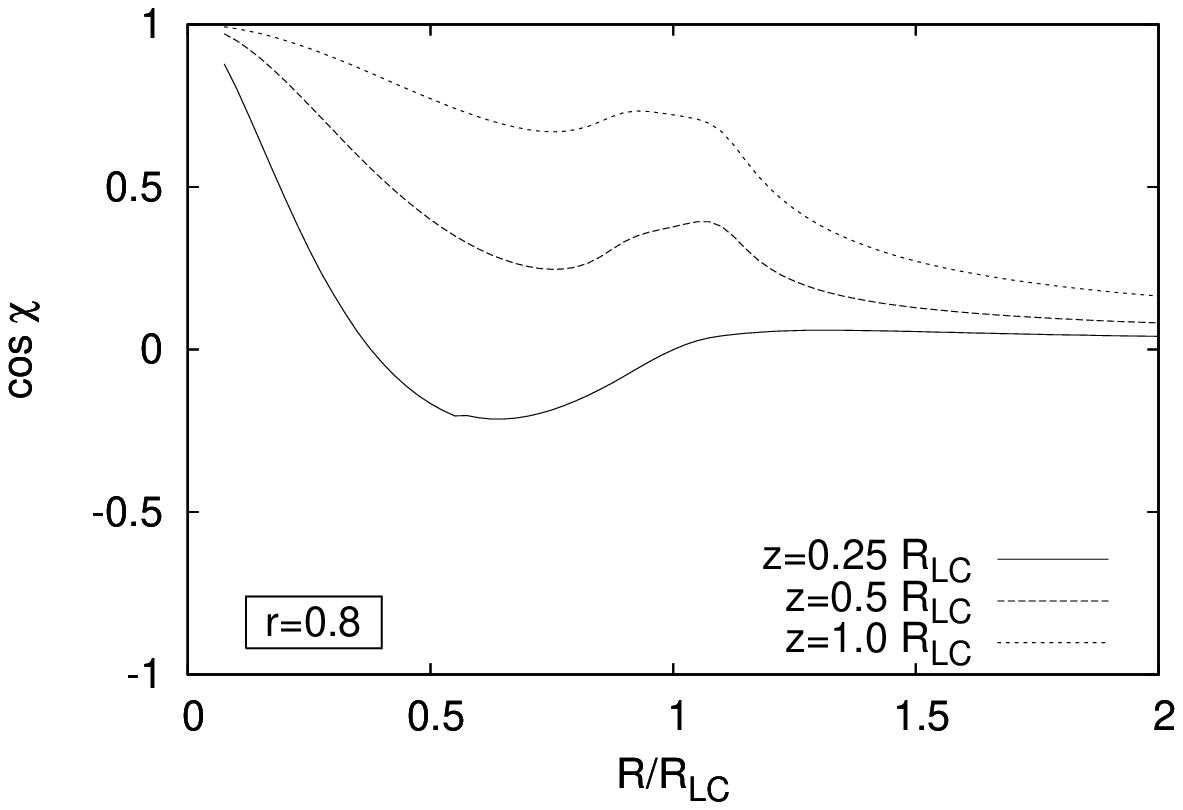}\\
 \includegraphics[width=63mm]{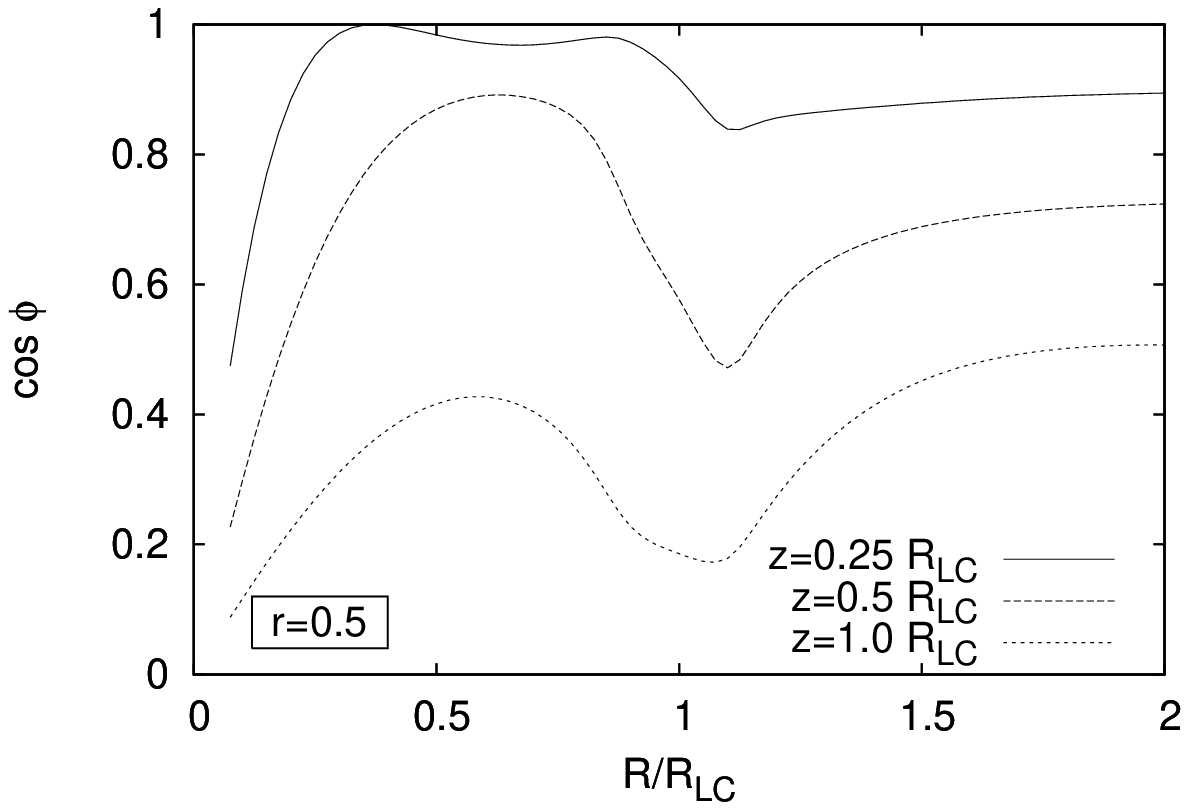}
 \includegraphics[width=63mm]{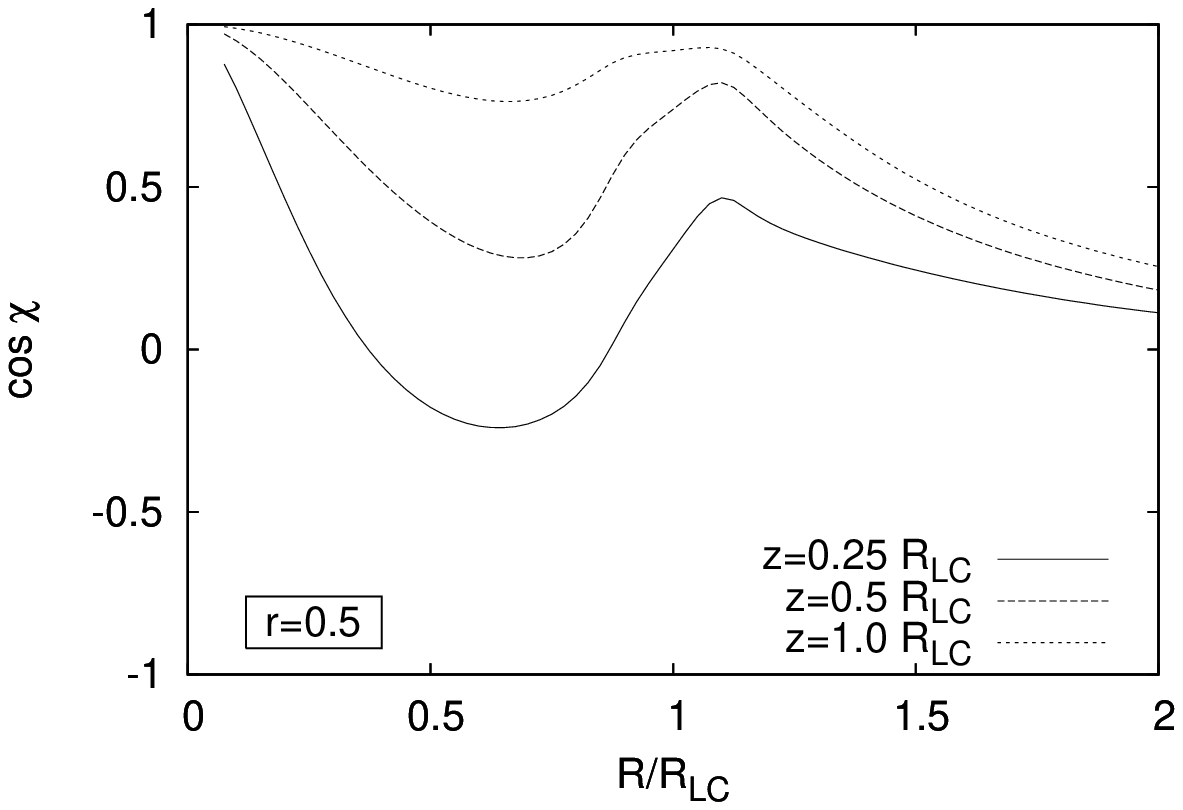}
\end{center}
 \caption{The angle between the magnetic field line and $R$ or $z$. 
 The top and bottom panels show the case with $r=0.8$ and $r=0.5$, respectively.
 We calculate the angle $\cos\phi$ and $\cos\chi$ with $z=0.25R_{\rm LC}$, 
 $0.5R_{\rm LC}$, and $1.0R_{\rm LC}$ for the both ratios.}
\label{fig:angle}
\end{figure*}

The difference of the magnetic field structure between the case with and without a current sheet 
suggests that the curvature radiation from the pulsar depends on whether a current sheet exists.
In the case with a current sheet, it seems that the curvature radiation from the magnetosphere would occur
uniformly in the whole region. On the other hand, in the case without a current sheet, the curvature radiation would
concentrate in a polar region. 
We show the angle between the magnetic field line and $R$ or $z$ axis in Fig.~\ref{fig:angle}.
These angles, $\phi$ and $\chi$, are given by 
\begin{eqnarray}
 \cos{\phi}&=&\frac{\vec{B}\cdot\vec{e}_{R}}{|\vec{B}|}, \\
 \cos{\chi}&=&\frac{\vec{B}\cdot\vec{e}_{z}}{|\vec{B}|}, 
\end{eqnarray}
where $\vec{e}_{R}$ and $\vec{e}_z$ are a unit vector tangent to the $R$ and $z$ direction, respectively.
The data are useful to calculate the curvature radiation, which is a plausible mechanism for observed 
pulsar radio emission \citep{1975ApJ...196...51R}. 

Our models do not satisfy the force-free condition near the light cylinder. However, in a
concerning realistic astrophysical situation, the force-free approximation would break at the fast-magnetosonic 
point which is inside the light cylinder. Therefore, our models could be good enough to be used in a in realistic astrophysical situation. 
Moreover, the physical properties discussed in this paper are the same even if we 
discuss the properties only inside the light cylinder.
It is needless to say that effect of inertia of plasma is important in the wind zone
and dissipation should also be concerned for photon emission (e.g., the 
outer gap model or slot gap model, which are strong around the last open field line). Therefore, another
treatment beyond the force-free approximation should be concerned, which is our future work.
The violation of the force-free condition in the vicinity of the light cylinder in our solutions means 
that they implicitly include dissipation around there. Such solutions with dissipation in the vicinity of the light cylinder 
have been already discussed in \citet{1994MNRAS.271..621M} in the context of magnetohydrodynamics.
Interestingly, their result is similar to our result in the case without a current sheet
(see figure 5 in their paper). Our solutions might express their situation and the appearance of the 
jet-like structure which is seen in our solution in the case without a current sheet might be an effect of 
inertia and dissipation. More research is needed to confirm that. 

In the framework with stationarity, we cannot answer if a magnetosphere without a current sheet appears and 
if the jet-like structure is realized. The Chandra X-ray satellite \citep{2000ApJ...536L..81W} has recently found the jet-like 
structure at the polar region of the Crab pulsar wind nebula.
Recent relativistic magnetohydrodynamical studies have shown that this jet can result from
compression by the magnetic hoop stresses or large scale vortexes 
\citep{2004A&A...421.1063D,2003MNRAS.344L..93K,2002MNRAS.329L..34L}. 
Although it is difficult to show that our jet-like structure is
responsible for the Crab jet due to the force-free approximation, we think that our results can be
a new probable initial condition for the RMHD studies of pulsar wind nebula, instead of the split
monopole model.
 
\bigskip
We would like to thank Prof. Y. Kojima, Prof. S. Shibata, and Dr. C.-M. Yoo for helpful 
discussion and comments. YT would like to thank Prof. V. S. Beskin, Prof. Ya. N. Istomin, and Dr. 
A. A. Philippov for giving meaningful comments on this work when YT visited Levedev Physical 
Institute. The visit was supported by the JSPS Institutional Program for Young Researcher Overseas 
Visits ``Promoting international young researchers in mathematics and mathematical sciences led by 
OCAMI''.  One of the authors (M.T.) is supported by the Postdoctoral Fellowships
for Research Abroad program by the Japan Society for the Promotion of
Science No. 20130253.
This work has been supported in part by the Grants-in-Aid for the Scientific Research Grant 
Number 23840023, 25103511 from the Ministry of Education, Culture, Sports, Science and Technology(MEXT) 
of Japan and HPCI Strategic Program of Japanese MEXT.

\appendix
\section{Derivation of the new trial toroidal current} \label{new-current}
In this appendix, we will derive $S_{\rm Tnew}$. 
First of all, let us derive $S_{\rm Tin}$, $S_{\rm Tout}$, and $S_{\rm TLC}$.
From Eqs.~(\ref{Ampere2}) and (\ref{FF2}), we have
\begin{eqnarray}
 \frac{R^2\Omega_{\rm F}^2}{c^2}S_{\rm T}(R,z)&-&\frac{2R\Omega_{\rm F}^2}{c^2}\partial_R\Psi
 \nonumber \\
&+&\frac{16\pi^2}{c^2}I_{\rm P}I_{\rm P}'=S_{\rm T}(R,z), \label{JT}
\end{eqnarray}
$S_{\rm T}(R,z):=8\pi^2 RJ_{\rm T}(R,z)/c$ in the right or left hand side of the above equation can be 
regarded as a trail toroidal current 
$S_{\rm Ttrial}(R,z)$.
Thus, we can make the following two equations
\footnote{To avoid the divergence at the axis $R=0$, we added $S_{\rm T}$ both sides of 
Eq.~(\ref{JT}) before making Eqs.~(\ref{JTin2}) and (\ref{JTout2}).}:
\begin{eqnarray}
 && S_{\rm Tin}(R,z)
 = \frac{1}{2}\left\{(1+R^2\Omega_{\rm F}^2/c^2)S_{\rm Ttrial}(R,z)\right.\nonumber \\
 &&~~~~~~~~~~~~~~~~~~~\left.-2\Omega_{\rm F}^2R\partial_R\Psi/c^2+16\pi^2I_{\rm P}I_{\rm P}'/c^2\right\}, \label{JTin2} \\
 && S_{\rm Tout}(R,z)
 =(1+R^2\Omega_{\rm F}^2/c^2)^{-1}\left\{2S_{\rm Ttrial}(R,z)\right. \nonumber \\
 &&~~~~~~~~~~~~~~~~~~~\left.+2\Omega_{\rm F}^2R\partial_R\Psi/c^2-16\pi^2I_{\rm P}I_{\rm P}'/c^2\right\},  \label{JTout2}
\end{eqnarray}
$S_{\rm Tin}$ or $S_{\rm Tout}$ can be regarded as the new toroidal current in our method. Unfortunately, the iteration 
only converges if the numerical domain is inside or outside the light cylinder. Then, in order to use 
our method in a domain including the light cylinder, by using $S_{\rm Tin}$, $S_{\rm Tout}$ and the hyperbolic tangent,
we make the following function $S_{\rm Tside}(R,z)$:
\begin{eqnarray}
 &&S_{\rm Tside}(R,z) \nonumber \\
 && =\frac{1+\tanh(\eta D)}{2}S_{\rm Tin}(R,z)+\frac{1-\tanh(\eta D)}{2}S_{\rm Tout}(R,z). 
 \nonumber \\
\end{eqnarray}
Thanks to the hyperbolic tangent, $S_{\rm Tside}$ almost becomes $S_{\rm Tin}$ and $S_{\rm Tout}$ inside and the outside the light 
cylinder, which is expected to make our iteration converge across the light cylinder.

$S_{\rm Tside}$ has useful form to make the iteration converge, however, we still need modification.
From Eqs.~(\ref{JTin2}) and (\ref{JTout2}), it is easy to see that both of $S_{\rm Tin}$ and $S_{\rm Tout}$ become $S_{\rm Ttrial}$ 
at the light cylinder, that is, $S_{\rm Tside}(R_{\rm LC},z)=S_{\rm Ttrial}(R_{\rm LC},z)$. Thus, it forces the value of $S_{\rm Tside}$ 
at the light cylinder to fix during the iteration.
We do not know the exact value of the toroidal current at the light cylinder a priori, and therefore
we need some modification to $S_{\rm Tside}$ in order to change the value at the light cylinder during the iteration.

Let us return to Eq.~(\ref{JT}). The exact form of the toroidal current is given by
\begin{equation}
 S_{\rm T}(R,z)
 =\frac{-2\Omega_{\rm F}^2R\partial_R\Psi/c^2+16\pi^2I_{\rm P}I_{\rm P}'/c^2}{\left(1-R^2\Omega_{\rm F}^2/c^2\right)}
 :=\frac{N}{D}
\end{equation}
Let the regularity condition at the light cylinder, $N=0$ at $R=R_{\rm LC}$, be satisfied.
Then, by using l'Hopital's rule, the toroidal current at the light cylinder is given by
\begin{eqnarray}
 \frac{N}{D}
 &&= \frac{\partial_R N}{\partial_R D}\bigg{|}_{R=R_{\rm LC}} \nonumber \\
 &&=\partial_R^2\Psi|_{R=R_{\rm LC}}
  +R_{\rm LC}^{-1}\Omega_{\rm F}^2\partial_R\Psi|_{R=R_{\rm LC}} \nonumber \\
 &&~~-\frac{1}{2}R_{\rm LC}^{-1}\Omega_{\rm F}^{-2}(16\pi^2I_{\rm P}I_{\rm P}')'\partial_R\Psi|_{R=R_{\rm LC}}, \label{N/D}
\end{eqnarray}
This is $S_{\rm TLC}(R_{\rm LC},z)$ and gives the exact value at the light cylinder formally.

Finally, by using the Gauss function, we have the following 
form of the new trial toroidal current:
\begin{eqnarray}
&&S_{\rm Tnew}(R,z) \nonumber \\
&&=\left\{\frac{1+\tanh(\eta D)}{2}S_{\rm Tin}(R,z)\right. \nonumber \\
&&\left.~~~~+\frac{1-\tanh(\eta D)}{2}S_{\rm Tout}(R,z)\right\}
     \left\{1-{\rm e}^{-D^2/(2\sigma^2)}\right\} \nonumber \\
	&&~~~~~ +S_{\rm TLC}(R_{\rm LC},z){\rm e}^{-D^2/(2\sigma^2)}, 
\end{eqnarray}
Thanks to the Gauss function, $S_{\rm Tnew}$ becomes $S_{\rm TLC}$ at the light cylinder.

\section{Criterion of the convergence}\label{convergency}
We discuss the criterion of the convergence of our calculation here.
During the iteration, we solve Ampere's law (\ref{Ampere2}) with various $S_{\rm T}$ 
that is an elliptic type differential equation.
Solving Eq.~(\ref{Ampere2}), we use the conjugate gradient method and the convergence obeys the method.
In addition, we should check whether $S_{\rm T}$ converges and it is determined as follows.

\begin{figure}
\begin{center}
\includegraphics[width=84mm]{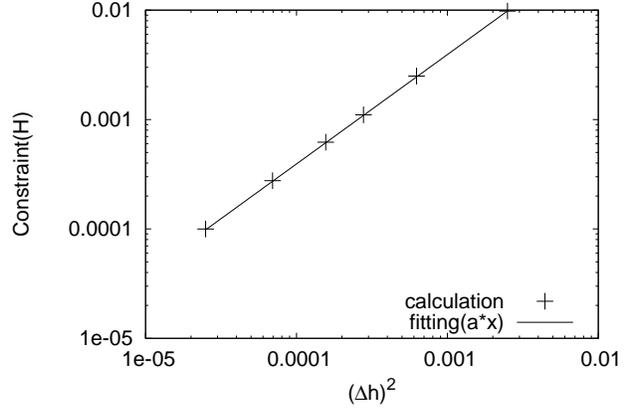} 
\end{center}
\caption{The plots of $H$ in the case that the outer boundaries are set 
at $R_{\rm max}=z_{\rm max}=2R_{\rm LC}$.
The vertical and the horizontal axis are $H$ and $\Delta h^2$, respectively. 
The criterion of the convergence we imposed are satisfied with the second-order accuracy.}
\label{hconst}
\end{figure}

Let us introduce the following quantity
\begin{equation}
 H_{P} = S_{\rm T}^{N+1}+\partial_{R}^2\Psi^{N}-\frac{1}{R}\partial_R\Psi^{N}+\partial_z^2\Psi^{N},
\end{equation}
where $N$ indicates the number of iteration times, and
$S^{N+1}_{\rm T}$ is determined from Eq.~(\ref{JTnew}) with $\Psi^{N}$.
We then integrate the magnitude of $H_{P}$ over the numerical domain, $\cal A$, as
\begin{equation}
 \frac{1}{S}\int_{\cal A}|H_P|\,dS\equiv H,
\end{equation}
where $S$ is the area of $\cal A$. We use $H$ to judge whether $S_{\rm T}$ converges. 
If $H$ can be numerically ignored, there is no difference between $S^{N+1}_{\rm T}$ 
and $S^{N}_{\rm T}$.
In our calculations, the derivatives of $\Psi$ are written
with the second-order accuracy. Therefore, we continue our routine until
$H$ can be ignored with the second-order accuracy. 
We show an example in Fig.~\ref{hconst} and see that our method that gives a next 
toroidal current density as Eq.~(\ref{JTnew}) shows the second order convergence at least.
In this calculation, we give $\eta$ and $\sigma$ in $S_{\rm Tnew}$
as $\sigma=4\Delta h$ and $\eta=0.5\sigma^{-2}$ where $\Delta h(=\Delta R=\Delta z)$ is the grid interval.
It could be expected that Eqs.~(\ref{Ampere2}) and (\ref{FF2}) are numerically satisfied simultaneously 
with $\Psi^{N}$ and $S^{N}_{\rm T}$ after $S_{\rm T}$ converges. However, it is not guarantee 
at the light cylinder because we do not impose the 
light cylinder condition during the iteration. Although whether the force-free condition is satisfied at light cylinder
is not guarantee in our method, our method gives a convergence solution of Ampere's law with
$S_{\rm T}$ giving by Eq.~(\ref{JTnew}).


\end{document}